\documentclass[onecolumn,authoryear]{els-mrw} 

\usepackage{amsmath,amssymb,amsfonts,amsthm,makeidx,graphicx}
\usepackage{txfonts}
\usepackage{helvet}
\newcommand{\araa}{\textit{Annual Reviews of Astronomy \& Astrophysics}}
\newcommand{\aj}{\textit{Astronomical Journal}}
\newcommand{\apj}{\textit{Astrophysical Journal}}
\newcommand{\apjs}{\textit{Astrophysical Journal, Supplement}}
\newcommand{\aap}{\textit{Astronomy and Astrophysics}}
\newcommand{\apjl}{\textit{Astrophysical Journal, Letters}}
\newcommand{\icarus}{\textit{Icarus}}
\newcommand{\mnras}{\textit{Monthly Notices of the RAS}}
\newcommand{\psj}{\textit{Planetary Science Journal}}
\newcommand{\nat}{\textit{Nature}}

\begin{document}

\chapter{Small Bodies in the Distant Solar System}

\author[1]{Samantha M. Lawler}%
\author[2]{Rosemary E. Pike}%

\address[1]{\orgname{University of Regina}, \orgdiv{Campion College and the Department of Physics}, \orgaddress{Regina, SK, Canada}}
\address[2]{\orgname{Center for Astrophysics $|$ Harvard \& Smithsonian}, \orgaddress{Cambridge, MA, USA}}

\articletag{Chapter Article tagline: Minor planets, asteroids, comets and interplanetary dust exterior to 30 au}

\maketitle

\begin{BoxTypeA}[chap1:box1]{Key points}
\begin{itemize}
\item The Kuiper Belt is made of leftover planetesimals from the planet formation era in our Solar System. 
\item Certain objects will be easier to discover and measure than others due to observation biases, but these biases can be quantified and accounted for to understand the real, unbiased distributions
\item The primary properties of these planetesimals that can be measured by telescopes, including dynamical properties (orbits) and physical properties (sizes, colors, composition, etc.). 
\item The sizes and shapes of Kuiper Belt objects tell us about the details of planet formation, while Kuiper Belt orbital distribution puts constraints exactly how and when the giant planets migrated.
\end{itemize}

\end{BoxTypeA}

\begin{glossary}[Glossary]
\term{absolute magnitude} Note that this term is used very differently for Solar System bodies than stars or galaxies. Abbreviated $H$, this is the magnitude an object would have as viewed from 1~AU distance, from the point of view of the Sun.

\term{albedo} A measure of reflectivity in a particular wavelength or range of wavelengths.  An albedo of 0 is a perfect absorber and reflects nothing, while an albedo of 1 is a perfect mirror and reflects all light received.

\term{apocentre} Abbreviated $Q$, the farthest point in an elliptical orbit to the central body.

\term{argument of pericentre} Abbreviated $\omega$, the rotational orientation of an orbit.

\term{astrometry} Precise measurements of the positions of objects - measuring a planet or small body's position on the sky over time allows an orbit to be computed.

\term{dwarf planet} A Solar System body that is large enough to pull itself into a spherical shape via self-gravity and hydrostatic equilibrium, but has not cleared its orbit of similarly sized bodies.

\term{eccentricity} Abbreviated $e$, a measure of the ellipticity of an orbit: a perfect circle has $e=0$, and $e$ values below $e=1$ represent closed orbits.

\term{ephermeris} The positions of a moving body over time, typically used for recovering a previously discovered moving object.

\term{inclination} Abbreviated $i$, a measure of the tilt of an orbit.

\term{longitude of ascending node} Abbreviated $\Omega$, 
the twist of an orbit.

\term{mean anomaly} Abbreviated $\mathcal{M}$, a time-averaged version of the true anomaly.

\term{Oort Cloud} Small icy bodies at very large distances from the Sun, ranging from the outer edge of the Kuiper Belt at $a\simeq2,000$~AU to the edge of the Sun's gravitational influence at $a\simeq100,000$~AU, these bodies are typically only observable only when they approach the Sun, sublimate, and become comets.

\term{opposition} The time during the year when an object in the sky rises close to sunset and sets close to sunrise, crossing the meridian close to midnight. Usually the best time of year to observe a particular object.

\term{pericentre} Abbreviated $q$, the closest point in an elliptical orbit to the central body.

\term{semimajor axis} Abbreviated $a$, half the distance between the pericentre and apocentre of an orbit.

\term{true anomaly} Abbreviated $f$, the angular position of an object around its orbit.

\end{glossary}

\begin{glossary}[Nomenclature]
\begin{tabular}{@{}lp{34pc}@{}}
AU &Astronomical Unit, the average distance between the Earth and the Sun, $1.5\times10^{11}$~m\\
MPC &Minor Planet Center\\
TNO &Trans-Neptunian Object\\
\end{tabular}
\end{glossary}

\begin{abstract}[Abstract]
The small bodies in the Kuiper Belt region of the distant Solar System are leftovers from planet formation.  Their orbital distribution today tells us about how giant planets migrated, while their surface properties, shapes, and sizes tell us about formation processes and collision rates. 
Probing these intrinsic properties requires a careful understanding of the observational biases that are a part of any telescopic survey that discovers small bodies.
While many of the details of giant planet migration are now understood due to careful comparison between de-biased discoveries in the Kuiper Belt and computational simulations, some discoveries have orbits that are still not easy to explain. Upcoming surveys such as the planned survey on Vera Rubin Observatory will help us to leverage the Kuiper Belt and refine our knowledge about the formation and dynamical history of our own Solar System.
\end{abstract}

\section{Introduction: Overview of the Kuiper Belt}\label{chap1:sec1}

The Kuiper Belt\footnote{The Kuiper Belt, or Edgeworth-Kuiper Belt, or Trans-Neptunian region, are all interchangeable names for the small bodies that orbit at semimajor axes that are the same or larger than Neptune's at $a=30.1$~AU and $a<2000$~AU, where the \textbf{Oort Cloud} begins.} was predicted to be a dynamically cold, flat disk of planetesimals out beyond the gas giant planets left over from the planet formation process \citep{Edgeworth1949,Kuiper1951}. 
Pluto, at first thought to be an additional planet similar in size to Neptune \citep{Tombaugh1997}, was decades later realized to be the largest (though not most massive) member of the Trans-Neptunian Object (TNO) population.
The second TNO was discovered in the early 1990s \citep{Jewitt1993}, and TNO discoveries ramped up very quickly with the advent of sensitive digital cameras, software algorithms to detect moving objects, and increasingly powerful computational resources.

The Kuiper Belt is full of fascinating individual objects, and we here highlight a few of the more unusual TNO discoveries over the past three decades (Table~\ref{tab:weird}).  The discovery of Eris \citep{Brown2005}, later measured to be more massive than Pluto \citep{Brown2007}, sparked the reclassification of Pluto from planet to a new small body class: a \textbf{dwarf planet}, large enough to pull itself into a sphere, but not large enough to clear its orbit of similarly-sized objects.  2008~KV$_{42}$ (528219), nicknamed ``Drac'', was the first TNO discovered to be moving backwards through the outer Solar System, with its orbit tilted 103$^{\circ}$ from Earth's orbit \citep{Gladman2009}.  Haumea is one of the largest TNOs, and is shaped like an American football \citep{Rabinowitz2006}, while having a ring \citep{Ortiz2017}, two moons \citep{Ragozzine2009}, and an associated group of smaller TNOs on orbits that can be traced back to a giant collision that broke them off of Haumea about one billion years ago \citep{Ragozzine2007}.  Sedna was the first high-pericenter TNO to be discovered \citep{Brown2004}, and its orbit remains hard to explain today \citep{Batygin2016,Bailey2019,Lykawka2023,Huang2024}.  
Additional observations of the first TNO, Pluto, have revealed fascinating details of its resonant orbit around the Sun \citep{Williams1971,Wan2001}, as well as the resonant orbits of all five of its moons \citep{Showalter2015}.  Arrokoth was the second \textit{New Horizons} flyby target, a tiny (30~km) TNO with a bizarre pancake-snowman shape \citep{Stern2019}, providing powerful evidence for the streaming instability (Section~\ref{sec:form}) as the dominant formation mechanism for small planetesimals.

\begin{table}[t]
\TBL{\caption{Notable TNOs}\label{tab:weird}}
{\begin{tabular*}{\textwidth}{@{\extracolsep{\fill}}@{}llll@{}}
\toprule
\multicolumn{1}{@{}l}{\TCH{TNO}} &
\multicolumn{1}{l}{\TCH{Notable for}} &
\multicolumn{1}{l}{\TCH{Diameter}} & \multicolumn{1}{l}{\TCH{Semimajor axis}} \\
\colrule
Pluto & Resonant orbit, 5 moons, largest TNO & 2,337~km & 39.5~AU\\
Eris & More massive than Pluto, inspired ``dwarf planet'' class & 2,326~km & 67.9~AU \\
``Drac'' & Highest inclination TNO ($i$~=~103$^{\circ}$) & 80~km & 42.0~AU \\
Haumea & Elongated shape, a ring, 2 moons & 2,100$\times$1,100~km & 43.1~AU\\
Sedna & Very high pericenter and large orbit & 900~km & 506~AU\\
Arrokoth & \textit{New Horizons} flyby target, contact binary & 30~km & 44.6~AU\\
\botrule
\end{tabular*}}{%
}%
\end{table}

\begin{figure}[t]
\centering
\includegraphics[width=.99\textwidth]{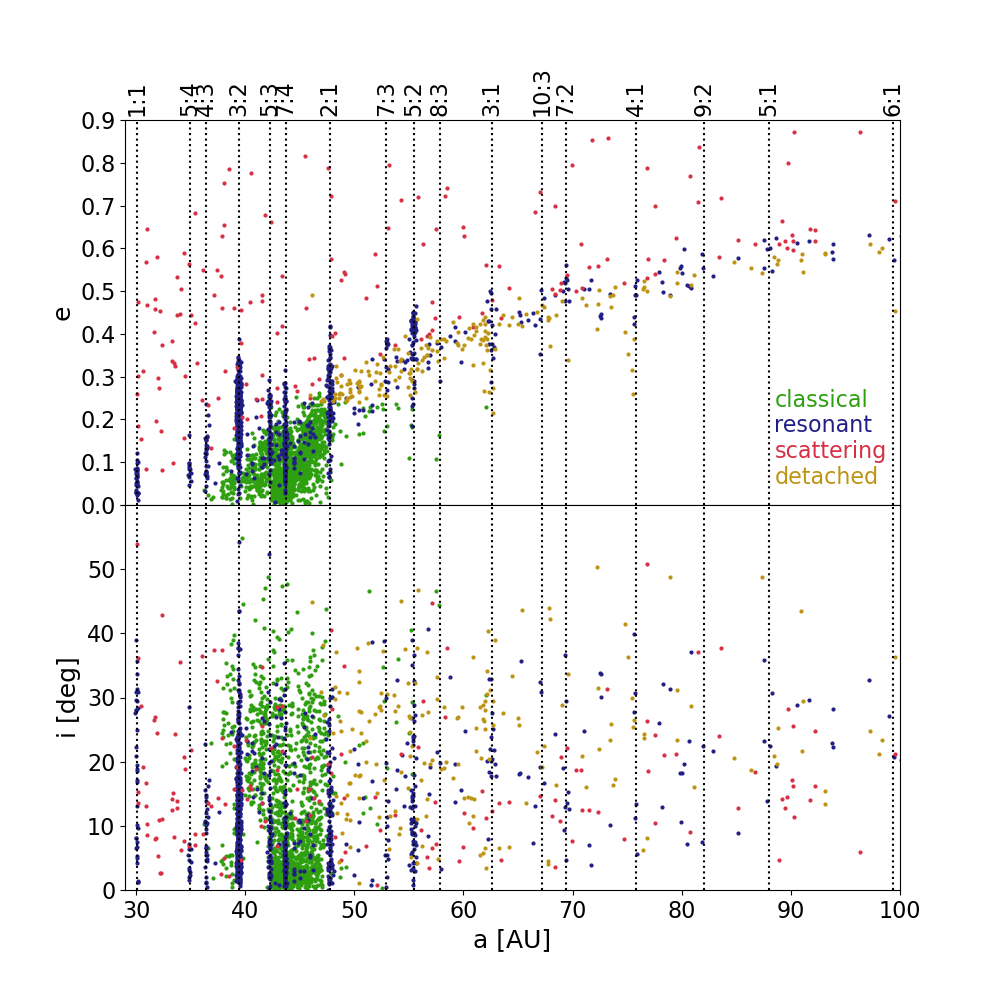}
\caption{Orbital elements of all multi-opposition TNOs (those with well-measured orbits) from the MPC Database as of December 2023.  Colors show the dynamical classifications from \citet{Volk2024} (discussed in more detail in Section~\ref{sec:orbclass}): green are classical TNOs, blue are resonant, red are scattering, and yellow are detached.  Dotted vertical lines show the semimajor axis where the orbital period ratio between a TNO and Neptune can be mathematically represented by small integers; these are the most significant resonances of Neptune.}
\label{fig:mpc}
\end{figure}

Over 3,000 TNOs are known today \citep{Volk2024}.
Collectively these objects are known as Trans-Neptunian Objects or Kuiper Belt Objects, and their orbital distributions teach us about the past formation and migration history of our Solar System. 
These objects occupy a large area of orbital space beyond Neptune.
Figure~\ref{fig:mpc} shows the orbital distribution of known TNOs in semimajor axis $a$, eccentricity $e$, and inclination $i$.  TNOs have semimajor axes ranging between that of Neptune ($a=30$~AU) and $a\simeq2000$~AU, where the Oort Cloud begins.   Most TNOs have semimajor axes between about 42-47~AU, with the densest part of the Kuiper Belt located between 43-44~AU.

\subsection{Measurable Properties of TNOs, and Chapter Outline}\label{sec:measurable}

Except for the TNOs Pluto and Arrokoth, which received flybys from the \textit{New Horizons} Mission \citep{Stern2020}, 
the extreme distance of the TNOs makes detailed observations challenging. When directly imaged, most TNOs appear only as point sources moving slowly relative to background stars.
However, we can gain a great deal of information by cleverly leveraging time and wavelength.  Most simply, the brightness of a TNO tells us information about a combination of its distance, its albedo (reflectivity), and its size.  These can be disentangled by making measurements of how the on-sky position changes over time, which gives the distance, and by measuring the TNO over a large wavelength range, which gives the albedo, discussed further in Section~\ref{sec:surf_props} \citep[e.g.,][]{Muller2009}.  
The distribution of TNO sizes is discussed in Section~\ref{sec:sizes}.

Some TNOs are bright enough (generally requiring them to be both large and close for ground-based observations) that spectroscopy can provide detailed information on chemical composition of the surface and atmospheric layers \citep[e.g.,][]{Pinilla-Alonso2020}.  Those that are not bright enough for spectroscopy can have colours measured in different photometric bands, effectively a very coarse spectrum \citep[e.g.][]{Pike2023}.  Surface properties are discussed in more detail in Section~\ref{sec:surf_props}.

Watching how a point source moves across the sky over the course of hours, months, and years gives information on its distance and orbit (discussed further in Section~\ref{sec:disc}).
Just as the motion of a body through space can be described with six cartesian position coordinates ($x$, $y$, $z$) and velocity coordinates ($v_x$, $v_y$, $v_z$), an orbit can be described in terms of six orbital elements.  Orbital elements are much more useful for talking about bodies within our Solar System, and the conversion between on-sky motion and orbits is handled through careful measurement and mathematical software, discussed in Section~\ref{sec:const_orbits}. For bodies orbiting the Sun in our Solar System, as TNOs are, the definitions of the six orbital elements are briefly defined below.  Much more detail on orbits and Solar System dynamics can be learned from \citet{Murray1999}.  The \textbf{semimajor axis}, $a$, defines the average distance of a body from the Sun.  The \textbf{eccentricity}, $e$, defines how elliptical an orbit is, with $e<1$ representing closed orbits, and $e=0$ being a perfectly circular orbit. Orbits that have $0<e<1$ will have differing distances from the Sun over the course of their orbits, with \textbf{pericenter} $q$ being the closest point to the Sun and \textbf{apocenter} $Q$ being the farthest.  \textbf{Inclination} $i$ describes the tilt of an orbit relative to the plane of Earth's orbit within the Solar System. 
The last three orbital elements help to place the orbit relative to a reference direction, which is chosen to be the position of the Sun on the celestial sphere on the Spring Equinox.  \textbf{Longitude of ascending node} $\Omega$ is the angle between the Spring Equinox direction and the point where the orbit crosses through the plane of the Solar System on its way north. \textbf{Argument of pericenter} $\omega$ is the angle between $\Omega$ and the pericenter position.  And lastly, \textbf{true anomaly} gives the angle between $\omega$ and the current position of the TNO on its orbit.

The Kuiper Belt is broadly divided into dynamical classes, which are identified depending on the behaviour of TNOs during long (typically at least 10~Myr) orbital simulations.  These dynamical classes are discussed in more detail in \citet{Gladman2008} and Section~\ref{sec:orbclass} below. 
These dynamical classes are useful for understanding how individual TNOs will behave on long timescales, as well as investigating how TNOs were emplaced onto the orbits where we observe them today. 

The Kuiper Belt functions as a fossil record of the past formation and migration history of the Solar System, particularly the giant planets.  Combining debiased observations (Section~\ref{sec:debias}) with dynamical classifications (Section~\ref{sec:orbclass}) and comparing with simulations of planetesimal formation (Section~\ref{sec:form}) and giant planet migration (Section~\ref{sec:dynamical_ev}) allows us to reconstruct the important details of how our planetary system formed and evolved.  The thousands of known exoplanets and dust disks around nearby stars in combination with our Kuiper Belt dust distribution (Section~\ref{sec:dust}) help us to place our Solar System into the broader context of planetary systems and illuminate the possible future evolution of our system \citep[e.g.][]{Li2022}. Upcoming surveys and targeted observations (Section~\ref{sec:future}) will help to answer some of the still-outstanding questions about how our Solar System achieved its current architecture.

\section{Discovering TNOs} \label{sec:disc}

All of the $>$3,000 TNOs now known were discovered in telescopic surveys of the sky.  TNOs are discovered through their motion relative to background stars.  Pluto was discovered by a human ``blinking'' between two photographs of the same spot on the sky taken hours apart, checking for anything that moved between the images.  Improved versions of this method are still used to discover TNOs, but now with the help of software.  

Two types of discovery methods are typically employed: individual image detections, like ``blinking'' described above, and ``shift and stack,'' where many images are taken, background stars are subtracted, and the images are added together after shifting at different rates and angles of motion \citep[e.g.][]{gladman1997,Fraser2024}.
Shift and stack processing, while much more computationally intensive, can detect TNOs fainter than the magnitude limit of individual exposures and thus discover smaller and more distant TNOs.

\begin{figure}[t]
\centering
\includegraphics[width=.99\textwidth]{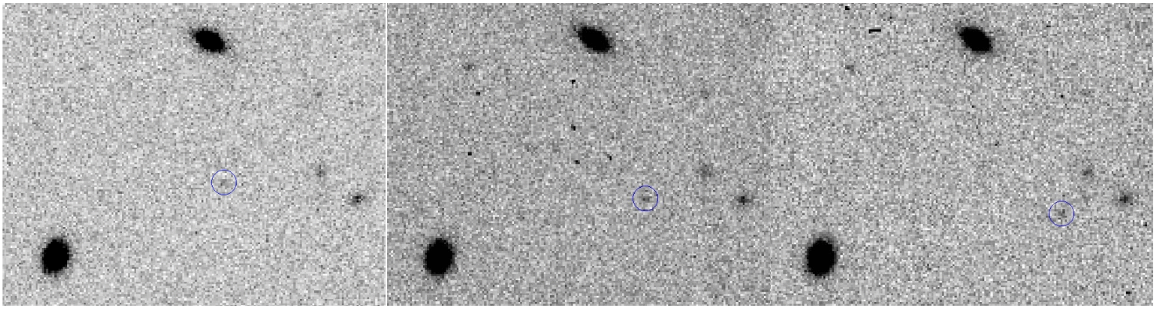}
\caption{Three images of the same patch of sky, with one TNO (circled in blue) moving visibly between the frames.  Each image is separated by about an hour in time.  Image from the OSSOS Survey/M.~Bannister. 
}
\label{fig:moving}
\end{figure}

To discover TNOs moving across individual images, software is used to create a catalog of stationary sources, and identify any detections which are not stationary (Figure~\ref{fig:moving}).
This can include moving objects, cosmic rays, transient sources such as supernovae, and other non-solar system objects.
Careful analysis is done to remove these contaminants, and then software is used to attempt to link possible moving object detections in different images together into possible on-sky motion for an object at a fixed distance \citep[e.g.][]{Petit2004}.  This is generally done with a single night of data, with as few as two or as many as dozens of images, though linking across several nights and even across months is possible \citep[e.g.][]{Holman2018,Rice2020}. 

Most TNOs have been discovered as part of large telescopic survey efforts, and below are descriptions of a few example recent and ongoing large TNO surveys (Table~\ref{tab:surveys}).  The Outer Solar System Origins Survey (OSSOS) detected the largest number of TNOs of any survey thus far: 840 TNOs with precisely-measured orbits, using the 4~m Canada-France-Hawaii Telescope (CFHT) on Maunakea to cover 155 square degrees on the sky between 2013-2018 \citep{Bannister2018}.  The Dark Energy Survey (DES) searched 5000 square degrees on the sky with the 4~m Cerro Blanco Telescope in Cerro Tololo between 2013-2019, and detected 812 TNOs \citep{Bernardinelli2022}.  The Large Inclination Distant Object Survey (L$i$DO) surveyed 34 square degrees centered at relatively high ecliptic latitudes, with the goal of finding more high-inclination TNOs.  L$i$DO discovered 140 TNOs (all with $i>15^{\circ}$) using CFHT between 2020-2023 \citep{Alexandersen2023}.  OSSOS, DES, and L$i$DO all detected TNOs by their motion between single images. The DECam Ecliptic Exploration Project (DEEP) also used the Cerro Blanco telescope, but between 2019-2023, and reaching much fainter detection limits by using shift and stack \citep{Trilling2024}. Analysis of their dataset is ongoing, but so far they have discovered over 100 TNOs in 60 square degrees of sky \citep{Smotherman2024}. The Classical and Large-a Solar System Survey (CLASSY) is surveying 10 square degrees, and is ongoing now (data collection planned 2022-2025), reaching much fainter detection depths than OSSOS by using shift and stack with the same telescope, CFHT \citep{Fraser2023CLASSY}.
In spite of the much smaller area of the CLASSY Survey, it is expected to find hundreds of TNOs because of the significantly fainter magnitude limit reached by stacking.
These surveys have discovered the majority of known TNOs.

\begin{table}[t]
\TBL{\caption{Selected Recent and Ongoing TNO Surveys}\label{tab:surveys}}
{\begin{tabular*}{\textwidth}{@{\extracolsep{\fill}}@{}lllll@{}}
\toprule
\multicolumn{1}{@{}l}{\TCH{Survey Name}} & \multicolumn{1}{@{}l}{\TCH{Years}} &
\multicolumn{1}{l}{\TCH{Telescope}} &
\multicolumn{1}{l}{\TCH{Sky coverage}} & \multicolumn{1}{l}{\TCH{TNOs discovered}} \\
\colrule
OSSOS & 2013-2018 & CFHT & 155~sq.~deg. & 840\\
DES & 2013-2019 & Cerro Blanco Telescope & 5,000~sq.~deg. & 812\\
DEEP & 2019-2023 & Cerro Blanco Telescope & 60~sq.~deg. & 100+ (in progress)\\
L$i$DO & 2020-2023 & CFHT & 34~sq.~deg. & 140\\
CLASSY & 2022-2025 & CFHT & 10~sq.~deg. & in progress\\
\botrule
\end{tabular*}}{%
}%
\end{table}

While the survey examples here have together detected nearly 2,000 TNOs, we expect that the Kuiper Belt contains several hundred thousand TNOs greater than 100~km in diameter \citep{Lawler2014,Crompvoets2022}.  As discussed in Section~\ref{sec:obsbias} below, every survey is biased toward preferentially detecting TNOs with certain properties. Thus, carefully measuring and publishing the properties of a survey (including brightness limits, positions on the sky, and tracking fraction), is important for understanding the real TNO populations that the discoveries were drawn from.

\subsection{Observation biases} \label{sec:obsbias}
TNOs are not detected uniformly by telescope observations - many effects lead to certain orbits, on-sky positions, and sizes being more easily detected than others.  In the early days of TNO searches, it was standard practice to publish just a list of TNO discoveries from a survey, but it is becoming increasingly common to publish the properties of a survey along with the discoveries \citep[e.g.][]{Bannister2018,Bernardinelli2024}.  Places on the sky where objects \textit{are not} discovered provides just as important a constraint as where objects \textit{are} discovered.

All of the biases within a survey can be modeled and accounted for by using a survey simulator \citep{Lawler2018}, if the survey is \textit{well-characterized}: pointing positions on-sky, observation dates, magnitude limits, and tracking rates are all measured and published.  The most likely TNOs to be discovered in a given survey are bright (which implies being large and having a large reflective area), close, and not moving either too quickly or too slowly across the sky.
Importantly, well-characterized surveys can help answer the question of whether the survey \textit{would} have detected objects with particular orbits \textit{if they existed}, or whether the survey was simply not sensitive to certain orbit types.

The dramatic biases in detections are primarily because TNOs are observed in reflected light, so the brightness is proportional to distance$^{-4}$, and thus we are extremely biased toward finding the closest TNOs.  This also means that TNOs on eccentric orbits are more likely to be detected than the same $a$ circular orbit, but only close to pericentre.  But each detection close to pericentre implies many more TNOs on similar orbits that were too faint to be detected because of their current distance.

Discovery biases can result in detections that are difficult to interpret unless a careful bias analysis is included.
Figure~\ref{fig:bias} shows an example simulation of how observational biases affect the likelihood of detection for a specific population with particular orbital element constraints and areas of higher on-sky density.  The black points in the face-down view show a snapshot of a model orbital distribution of Plutinos (TNOs in the 3:2 resonance with Neptune, like Pluto; see also Section~\ref{sec:res}).  The red points show simulated detections, that is, model objects that would have been detected by a survey that covered two areas on the sky (and thus, two directions through the Solar System) to a given magnitude limit. 
The normalized histograms on the left side of Figure~\ref{fig:bias} show the difference between selected orbital element distributions of the model population and the simulated detected population.  While the model population follows a standard inclination distribution \citep{Brown2001}, 
the detected population has two distinct peaks caused by the ecliptic latitudes chosen by each of the two survey blocks on the sky.  One was close to the ecliptic, and thus preferentially detected low-$i$ objects, while one was at a higher ecliptic latitude, and detected only higher-$i$ objects.  The biases in the longitude of ascending node $\Omega$ are similarly discrepant - while the model population is isotropic in $\Omega$, the simulated detections have two peaks, where the nodal crossing angle makes it most likely that the objects will be within the survey block on-sky.
Thus the apparent discrepancy between the detections and the model can likely be attributed only to detection biases, not issues with the model distribution.

We note that the recent proliferation of commercial satellites in low Earth orbit \citep[e.g.][]{Lawler2022,Bassa2022} have added new observing biases to all of astronomy, particularly to wide-field TNO surveys, and these biases are expected to grow worse as more satellites are launched.  Satellites reflect sunlight long into dark observing time, and typical Starlink satellites (which currently\footnote{As of 23 Aug 2024, there are 6,325 Starlink satellites in orbit (\url{https://planet4589.org/space/con/conlist.html}), out of a total 10,115 active satellites (\url{https://celestrak.org/NORAD/elements/}).} comprise the majority of satellites in orbit) are millions of times brighter than typical TNOs when sunlit.  Satellites fly through the field of view of long time exposure images and make bright lines that destroy any data near them.  The brightness and frequency of satellite streaks in a given image depends on the time of night, time of year, and exact on-sky observing position.  But satellite streaks will typically be worse for ecliptic observing in the local summertime, making TNOs and other solar system small bodies in that part of the Solar System less likely to be discovered. 

\subsection{Debiasing survey discoveries} \label{sec:debias}

The biases in orbital element distributions introduced by surveys are significant, but if the biases of a survey are all well-measured and published, those biases can be taken into account and the survey discoveries effectively \textit{de-biased}.  More details of the de-biasing process and science output examples can be learned from e.g., \citet{Lawler2018} about survey simulator software, but the basic concept is outlined here.

\begin{figure}[t]
\centering
\includegraphics[width=.99\textwidth]{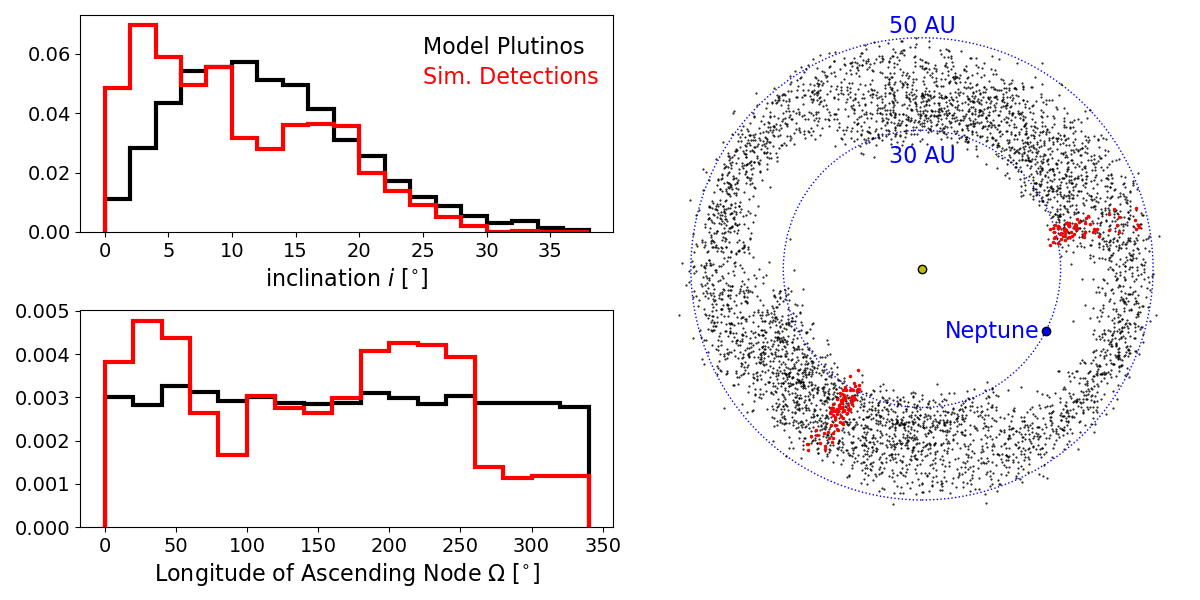}
\caption{Simulated detections in a model produced by the OSSOS Survey Simulator software, demonstrating observing biases.  The right plot shows a face-down view of a model TNO population (in this case, the Plutinos; black points), with simulated detections from a survey with two pointings on the sky shown in red points.  On the left, two histograms (where the y-axis shows relative fraction of objects per bin) show the significant difference in the distribution of orbital elements between the model (black) and simulated detections (red) in inclination $i$ (top left plot) and longitude of ascending node $\Omega$ (bottom left plot). Note that the apparent `bad fit' of the model to the simulated detections is purely the result of the observational biases of the survey.}
\label{fig:bias}
\end{figure}

A survey simulator is software that applies the biases of a particular survey or surveys to a set of model objects, to determine which model objects were in the right portion of the sky, bright enough, and moving at the right speed to be likely to have been detected by that survey.  For example, the OSSOS Survey Simulator is publicly available at \url{https://github.com/OSSOS/SurveySimulator}, the DES Survey Simulator is publicly available at \url{https://github.com/gbernstein/pixmappy}, and the Vera Rubin Observatory Survey Simulator is available at \url{https://github.com/dirac-institute/sorcha}.  In the example shown in Figure~\ref{fig:bias}, the biases from two observing blocks of the OSSOS survey were applied to a simulated orbital distribution of Plutinos using the OSSOS Survey Simulator \citep{Petit2018}.
Running a model through the survey simulator effectively biases the model to match the real TNO detections (which have the same observational biases), and then the biased model can be compared to the real detections directly.
These comparisons are usually done with statistical tests \citep[e.g.][]{Anderson1952}, which quantify the difference between the detections and the biased model.
An important subtlety of this approach is that more than one intrinsic model can produce biased detections that are not statistically rejectable by the real objects, particularly for populations with small numbers of discovered objects.
It is important to properly consider the complexity of the input model, and determine whether any additional complexity (for example, not using a uniform distribution) is merited, to avoid over-fitting.


\section{Formation and Evolution}

The planet formation process happened in our Solar System 4.5~Gyr ago, when no one was around to watch.  So we have to be clever by studying other, nearby protoplanetary systems that likely follow the same processes that took place in our own Solar System.  We can combine those extrasolar observations with other clues left behind in the distribution and composition of the small bodies in our own Solar System today.

\subsection{From Disk to Planetary System} \label{sec:form}

By measuring dust and gas in protoplanetary systems \citep{Andrews2020}, 
as well as via detailed computational simulations, we have learned the basic steps of planet formation.  Protoplanetary disks are initially composed of gas and very small dust grains.  This dust grows from micron-sized grains to cm-sized pebbles through electrostatic forces \citep{Steinpilz2020}.  Once the particles reach the size of pebbles, they become strongly affected by the motion of the gas in which they are embedded.  The ``streaming instability'' happens as pebble-sized particles embedded in gas are forced together by gas drag slowing down the leading particles and allowing those following behind to catch up and cause clumps \citep{Youdin2002}.  The streaming instability is what allows pebble cloud collapse to occur, which very effectively and quickly turns large clumps of pebbles into planetesimals: simulations show that a 100~km planetesimal can form directly from pebbles in only $\sim$100 years via this mechanism \citep{Robinson2020}.  These 100~km planetesimals are then large enough to grow into planets via gravitational focusing of smaller bodies within the disk.  The planetesimals that we see today in the Kuiper Belt are the ones that were never grew into larger bodies themselves, and also avoided being incorporated into planets.

Numerical simulations of the streaming instability and pebble cloud collapse make a tidy, testable prediction: nearly every planetesimal formed should originally be in a binary pair \citep{Nesvorny2021}.  As we discuss in Section~\ref{sec:classical}, this prediction appears to be true, supported by both the high proportion of measured binary TNOs in the cold classical sub-class in particular \citep{Fraser2017}, and in the shape of the \textit{New Horizons} Mission flyby target Arrokoth \citep{McKinnon2020}, 
which is a clear contact binary.

\subsection{Dynamical evolution} \label{sec:dynamical_ev}

The oddly eccentric and tilted orbit of the very first TNO, Pluto, was the first hint that the Kuiper Belt holds clues to the dynamical history of the Solar System.  \citet{Malhotra1993} first showed that outward migration of Neptune could have captured Pluto into the 3:2 mean-motion resonance as it swept outward, and this would naturally bump up Pluto's $e$ in the process.  However, the overall orbital distribution produced in simulations where Pluto sweeps outward through a disk of planetesimals does not do a great job of reproducing the orbits we observe, particularly the orbital distributions of TNOs in Neptune's mean-motion resonances \citep{Hahn2005}.

Another way to explain Pluto's weird orbit is through a different, much more chaotic and violent pathway to migration often called ``the Nice Model,'' where a dynamical instability between the giant planets causes Neptune to be flung outward onto an eccentric orbit that damps down as Neptune scatters and captures objects from the primordial Kuiper Belt \citep{Tsiganis2005}.  Some more recent modifications to the original Nice Model make it better reproduce the orbits that we see today.  Adding in an extra planet that is later ejected helps to keep the classical TNOs in the right place during giant planet migration \citep{Batygin2012}.  Making Neptune's migration ``grainy'' to simulate the scattering of the largest, Pluto-sized primordial TNOs that would have had enough mass to cause Neptune to have small jumps in $a$ while scattered helps to match the resonant and near-resonant orbital distributions \citep{Nesvorny2016}.  Giving Neptune one larger jump in $a$ helps to create a dense patch of orbits that we observe within the classical belt, called the ``kernel'' \citep{Nesvorny2015}.  With these additional modifications, the Nice Model appears to reproduce most of the TNO orbits we observe today.

Dynamical models must also create the Oort Cloud, which is comprised of planetesimals from the giant planet region that were ejected onto such large-$a$ orbits that they are subject to the isotropizing and pericenter-lifting forces of Galactic tides \citep{Dones2004}.  The violence of the Nice Model does an excellent job of ejecting large masses of small planetesimals onto orbits consistent with the Oort Cloud comet orbital distribution we observe today.
The orbital features within the Kuiper Belt all tell detailed stories about exactly how the giant planets reshuffled when the Solar System was still very young.

\section{Understanding the orbital and physical properties of TNOs}

The orbital distribution and physical properties of TNOs are the result of their formation and evolution.
The spatial and orbital distribution of TNOs is indicative of their migration history, and thus constrains migration of the Solar System's giant planets \citep[e.g.,][]{Levison2003,Nesvorny2015}.
The physical properties of TNOs include their composition, surface color, albedo, rotation rate, shape, the presence of satellites/binary companions, and sizes.
Composition, surface color, and albedo are all closely linked and typically measured using multi-band photometry or spectroscopy and likely reflect the composition of the regions of the proto-planetary disk from which the small bodies formed \citep[e.g.,][see Section~\ref{sec:surf_props}]{Fraser2012,Buchanan2022}.
The rotation rate, shape, binarity, presence of satellites, and size are measured using photometry and occultations, and these properties are indicative of how the objects formed and any collisional evolution they experienced during the lifetime of the solar system \citep[e.g.][]{Benecchi2013, Buie2020}.
The investigation of these physical properties thus provides insight into the formation and evolution of the solar system.

\subsection{Constraining and Computing Orbits} \label{sec:const_orbits}

Discovering TNOs is just the first step in understanding the distant solar system.
In order to understand the significance of each object and how it fits within the outer solar system, we need to measure accurate orbits for the objects.
Each astrometric measurement provide a location on-sky for the object, comparable to an $x$, $y$, and $z$ position.
The change in location over time between astrometric points gives us an instantaneous velocity, $v_x$, $v_y$, $v_z$.
It is possible to use these 6 coordinates to describe the object, but it is more informative to convert these 6 coordinates into Keplerian orbital elements with $a$, $e$, $i$, $\Omega$, $\omega$, and $\mathcal{M}$ as described in Section \ref{sec:measurable}.
The mean anomaly $\mathcal{M}$ is often used instead of the true anomaly.
$\mathcal{M}$ is a fictitious angle, which indicates the angular distance an object would have traveled along its orbit from perihelion if its rate of motion was constant, a time-averaged version of the true anomaly.
There are several orbital characteristics computed from these coordinates which are very useful for understanding orbits, including pericenter ($q$, closest sun approach), apocenter ($Q$, furthest sun distance), and period ($P$, the duration of a complete orbit).
TNO orbits are typically computed around the Solar System barycenter, with inclinations determined with respect to the ecliptic plane, the plane of Earth's orbit around the Sun.
The ecliptic plane is not \textit{exactly} aligned with the plane of the Kuiper belt (basically, the average $z$ position of all Kuiper Belt objects at each distance), which is closer to the invariable plane \citep[the averaged plane of the whole Solar System, taking into account the masses and inclinations of all the planets;][]{Souami2012,Volk2017}, 
however, this is typically sufficiently accurate for most analyses.
Computing orbital elements with high accuracy is the main goal of acquiring the astrometric measurements of TNOs.

Measuring a sufficient fraction of the orbit is required to compute the objects' orbits accurately.
Because TNOs are at such large distances from the Sun, with semi-major axes $a>30$~au, their orbital periods are hundreds of years.
A few measurements within a single night can only predict the on-sky position of that TNO accurately enough to unambiguously find it again within a few days.
In order to constrain the orbit of each TNO, multiple measurements of the TNO's position (astrometry) are needed.
As the overwhelming majority of TNOs are unresolved, even with space-based telescopes, conclusively linking different TNO astrometric measurements is a complicated task. 
Determining correct linkages between observations on different nights relies on fitting a straight line (in short arcs) or a Keplerian orbit (in longer arcs) to the position measurements and evaluating the residuals between the astrometry and the orbital fit.
Each astrometric point will have an associated uncertainty, which can range from $\sim0.05-1$ arcseconds depending on the number of stars in the frame with well measured positions.
With the new Gaia catalogs, astrometric precision has increased dramatically \citep{GaiaCollaboration2016}, which has been extremely important for improving the accuracy of moving object measurements.
The goal for tracked TNOs is typically several astrometric measurements, dispersed in time, and spanning at least two years.
This multi-year arc is required in order to determine an orbit with a semi-major axis uncertainty of fractions of an AU and to determine the object's orbital classification.

Significant effort has been invested into calculating orbits from astrometry.
Particularly for short arcs, this can involve simplifying assumptions.
Some of these include circular or nearly circular orbits and the smallest inclination necessary to provide a reasonable orbit fit \citep{Bernstein2000}.
As most TNOs do in fact have low-$i$ and low-$e$ orbits with $42<a<47$~au, these assumptions do provide a likely orbit fit.
Unfortunately, for anything outside the classical TNOs, these assumptions may provide inaccurate predictions of the objects' future positions on-sky, and result in lost discoveries preferentially for unusual orbits \citep{kavelaars2008}.
Multiple orbit fitting tools have been created, including the \citet{Bernstein2000} orbit fitting code, which is optimized for outer Solar System objects.
Orbit fitting code attempts to minimize the residuals between the astrometric measurements and the orbital fit, and also generate a prediction of position of the TNO at different times (called an \textbf{ephemeris}).
The ephemeris will have an associated uncertainty, which is typically much larger along the direction of the TNO's motion, as small uncertainties in the velocity of the measurements can translate to larger on-sky uncertainties.
The ephemerides are used to plan additional recovery observations, with the goal of gaining additional astrometric measurements to determine the TNO's orbit with higher and higher accuracy.
As the arc length over which astrometry is measured increases (assuming the astrometric quality is good), the uncertainty of both the orbit fit and the ephemeris decreases.

\begin{figure}[t]
\centering
\includegraphics[width=.5\textwidth]{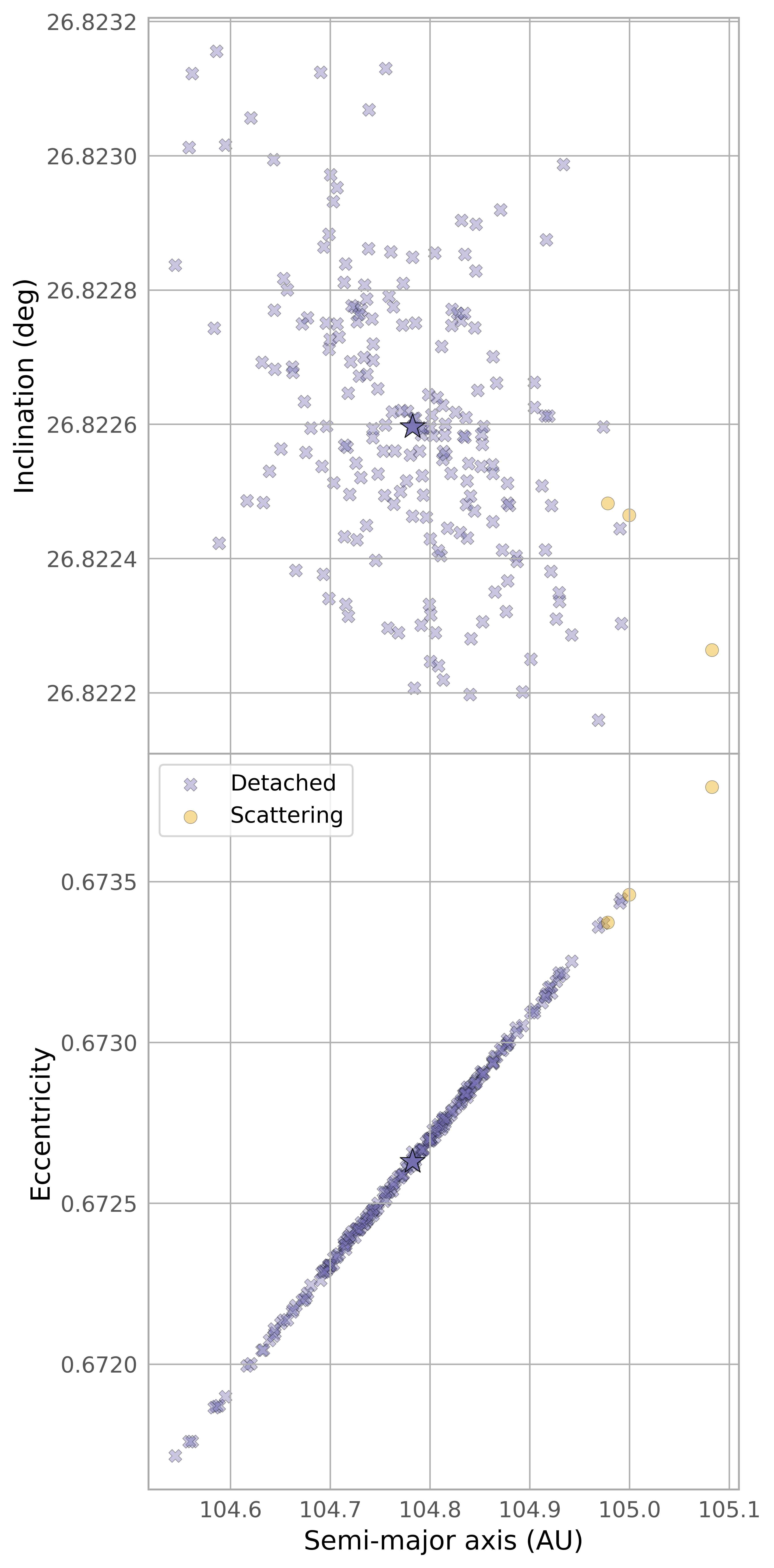}
\caption{The $a$, $e$, $i$ values for variant orbits of a real TNO from the L$i$DO Survey.  Each of these points shows an orbital fit which is consistent with the astrometry acquired for that TNO. The majority of the clones for this TNO have detached orbits, but a few are scattering (see Section \ref{sec:orbclass} for a description of the classes).  The discrepant classifications are at the largest-$e$ and $a$ values, and are likely not representative of the real TNO's behavior, as the overwhelming majority of the variants agree with each other. The interdependence of $a$ and $e$ in the fit is also evident in the bottom panel.  The `star' indicates the nominal or unfuzzed orbital fit.}
\label{fig:orbit_fitting}
\end{figure}

How many astrometric measurements are required for good orbits depends on the orbital accuracy needed, the cadence of measurements, and the arc length.  A typical astrometric sequence is 3-5 measurements in one night, with 1 or more additional measurements within a few nights, another measurement within a month, and another set of one or more measurements in each of the following two years.  Once enough astrometry is acquired, orbital fitting software can be used to determine the TNO's orbit.
Orbit fitting software will produce a fit which minimizes discrepancies between the measured astrometry and the predicted position based on the orbit.
This is referred to as the `nominal' orbit.
The nominal orbit can be extremely useful for on-sky predictions and for understanding the likely classification of the object, however, quantifying the uncertainty is also an important part of determining the orbit.
Orbit fitting software typically produces uncertainties in the orbital elements, which can be useful for a basic check of the orbital accuracy.
To replicate the interdependence of the orbital elements, however, a more effective way to determine the variety of orbits which are formally consistent with the measured astrometry is to re-sample the astrometric data.
To do this, we use a typical uncertainty for measuring astrometry (based on the telescope and star catalog used) and ``fuzz'' the astrometry by randomly drawing offset values based on that uncertainty.
The orbit fit is then repeated, and when this is done tens to hundreds of times, a cloud of variant orbits is produced, as shown in Figure \ref{fig:orbit_fitting}.
All of these orbits are consistent with the measured astrometry, and whether they are significantly different from each other is a major factor in deciding whether additional astrometry of the object is needed.

\subsection{Orbital Distribution and Orbit Classifications} \label{sec:orbclass}

\begin{figure}[t]
\centering
\includegraphics[width=.8\textwidth]{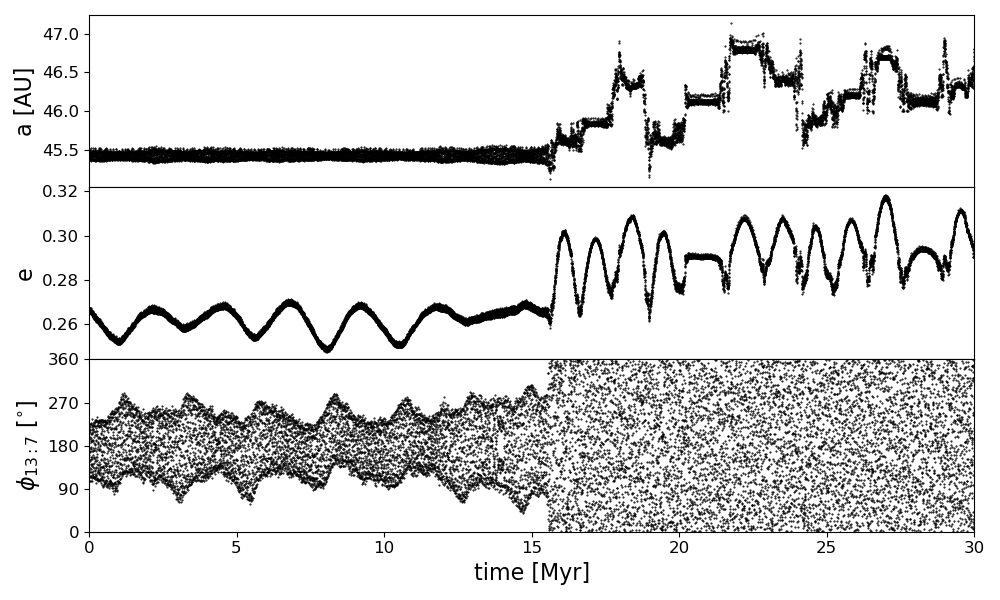}
\caption{Example orbital integration of one variant clone from a real TNO in the L$i$DO Survey \citep{Alexandersen2023}.  This particular variant experiences libration of $\phi_{13:7}$ for a few Myr, then falls out of the resonance and begins scattering.  When this happens, $a$ jumps by a large amount, and the resonant angle $\phi_{13:7}$ (see Eq.~\ref{eq:res}) switches to circulation rather than libration, taking on all values 0-360$^{\circ}$. This is an example of how TNOs can jump between classifications over the course of a simulation.}
\label{fig:integration}
\end{figure}

When an orbit is well-measured, the TNOs can be subdivided into different orbital classifications, which is helpful for determining how the TNOs were emplaced onto the orbits we observe them on today, their past history, and some information about their formation location.  As was nicely laid out in \citet{Gladman2008}, the key to dynamical classifications is to take the current, measured orbit and integrate forward under the influence of gravity for at least 10~Myr (we find at least 30~Myr is better if the sample includes TNOs with $a>100$~AU, or if you are trying to diagnose sub-resonances).  Figure~\ref{fig:integration} shows an example test particle integration using the open-source N-body code REBOUND\footnote{Excellent documentation, installation instructions, and examples available at \url{https://rebound.readthedocs.io/}} WHFAST \citep{Rein2015}.  Each row shows the behaviour of a different orbital element or orbital element combination for this test particle during the 30~Myr integration, which includes the test particle, as well as the four giant planets, all started on their true orbital positions (as obtained extremely accurately from the NASA Horizons Database\footnote{\url{https://ssd.jpl.nasa.gov/horizons/app.html#/}} for the same starting date).  This particular particle experiences stable resonance for about 15~Myr (see Section~\ref{sec:res} below) with a consistent value of $a$ and slowly oscillating $e$.  After this point, the particle leaves the resonance and $a$ changes rapidly due to stronger gravitational interactions with Neptune.  The particle in Figure~\ref{fig:integration} experiences two different dynamical classes during the integration: it is resonant for the first half of the integration, and scattering for the second half.  

Each real discovered TNO has uncertainties on-sky, that result in uncertainties in the orbital elements, and each variant orbit (Figure~\ref{fig:orbit_fitting}) consistent with the astrometry needs to be integrated and classified.  Then classification criteria are needed: how many variants need to stay inside a particular class for what fraction of the integration?  Different surveys use different criteria.  The L$i$DO Survey, for example, requires at least 80\% of clones to  be in the same classification for at least 80\% of the integration in order to consider a TNO secure in a particular class, otherwise it is classified as ``insecure'' in whichever classification it spends most time.  These ``insecure'' classified TNOs can eventually be upgraded to a secure classification with additional astrometric measurements. 

\begin{figure}[t]
\centering
\includegraphics[width=.7\textwidth]{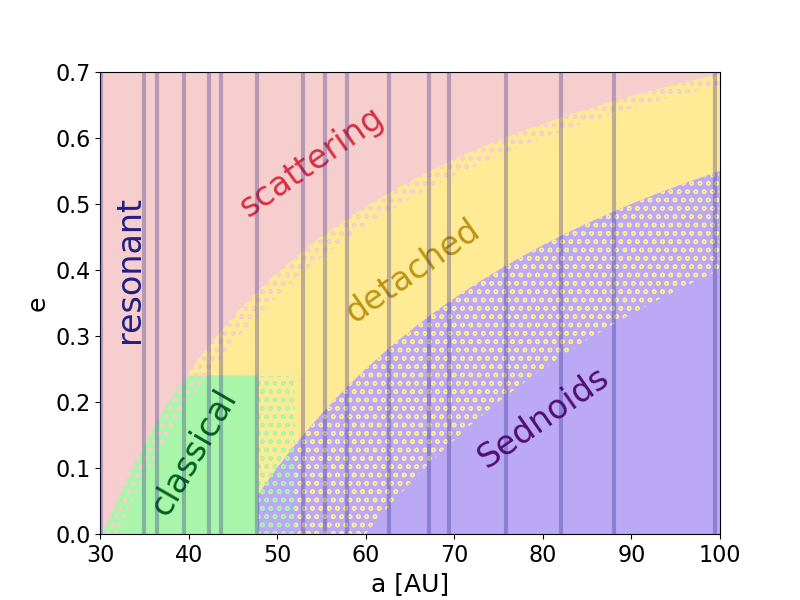}
\caption{A rough schematic of dynamical classes in the Kuiper Belt based on $a$ and $e$, with hatched areas showing where significant overlap typically happens between classes (some of this overlap is visible in the classifications for real TNOs in Figure~\ref{fig:mpc}), or where boundaries are not well defined between classes in $a$-$e$ space.  As discussed in the text, this diagram alone cannot be used for dynamical classification, it just provides guidelines.  Orbital integrations of at least 10~Myr in duration are required for dynamical classification.}
\label{fig:classify}
\end{figure}

Figure~\ref{fig:classify} shows a schematic of the Kuiper Belt dynamical classes, but this is only a rough guide and cannot be used for classification (as hinted at by the hatched regions between some of the classes); the only way to truly determine orbital class is by using orbital integrations as described above.  Below, we describe the different dynamical classes as originally outlined in \citet{Gladman2008} plus an additional class now required due to much more distant TNOs that have been discovered in the past decade and a half.

\subsubsection{Centaurs and Oort Cloud: Outside of the Kuiper Belt}

Objects in the outer Solar System with $a<30$~AU or $a>2000$~AU are not part of the Kuiper Belt.  Objects that have $a<30$~AU are Centaurs, which experience relatively frequent, strong gravitational interactions with the giant planets and are likely to be ejected from the Solar System on few Myr timescales \citep{Dones1996}.  Objects with $a>2000$~AU experience significant tidal perturbations from the Milky Way's mass distribution, causing oscillations in pericenter and other orbital elements over time, resulting in an isotropic inclination distribution, and are considered to be part of the Oort Cloud \citep{Dones2004,Veras2014}. 

\subsubsection{Scattering TNOs}

The scattering TNOs come close to Neptune's orbit and experience large gravitational perturbations that cause large changes in semimajor axis on the timescale of at most a few million years.  If an orbital integration displays a change in $a$ of more than 1.5~AU in 10~Myr, the TNO is scattering.  These TNOs are being slowly depleted from an original population that was approximately 100 times the current mass, emplaced during Neptune's migration phase \citep{Duncan1997}.  There is a small amount of resupply as TNOs leak out of resonances and the Oort Cloud, but this population is overall declining over time.

\subsubsection{Resonant TNOs} \label{sec:res}

Resonant TNOs have orbital periods that are close to small integer ratios with Neptune's orbital period.  Pluto, for example, is in the 3:2 resonance with Neptune: Neptune completes 3 full orbits of the Sun in the time that Pluto completes exactly 2 full orbits.  But not all TNOs meeting this rather loose criteria will be resonant; they must also exhibit a particular angular confinement.  The resonant angle $\phi_{j:k}$ is defined for the $j$:$k$ resonance as a combination of the orbital angles of Neptune and the TNO (subscript N and TNO in Equation~\ref{eq:res}, respectively), with $\lambda=\Omega+\omega+\mathcal{M}$:

\begin{align}
\phi_{j:k}=j\lambda_{\rm TNO} - k\lambda_{\rm N} - (j-k) (\Omega_{\rm TNO}+\omega_{\rm TNO})
\end{align} \label{eq:res}

When $\phi_{j:k}$ is confined (this means it is librating, not circulating, as in the first half of the integration shown in Figure~\ref{fig:integration}) over the course of an orbital integration, that means that the combination of orbital angles is actually protecting the TNO from close encounters with Neptune\footnote{Supplemental video 2 in \citet{Gladman2021} nicely visualizes how the confinement of $\phi_{j:k}$ protects resonant TNOs from close encounters with Neptune over many orbits and many $\phi_{j:k}$ libration cycles.}, and keeping the TNO coupled to Neptune's orbit with gravitational kicks always happening near the same orbital angles, similar to pushing a kid on a swing.  The $j$:$k$ resonance that is most likely to dominate is easy to calculate from the period ratio between the TNO and Neptune, and confinement of the resonant angle $\phi_{j:k}$ can be searched for numerically.  There are also small, fast ($<1000$~yr) periodic oscillations in $a$ caused by the resonant condition, and often other orbital elements show correlated or near correlated oscillations as well.  The combination of all of these can be used to diagnose resonance algorithmically, or using machine learning strategies \citep{Volk2024b}. 

Resonant TNOs are particularly useful for several reasons.  The protection from close encounters with Neptune means that they have on-average lower pericenter distances than other stable populations at the same $a$, making them easier to observe.  Due to the confinement of orbital angles by $\phi_{j:k}$, they come to pericenter at certain places on the sky, making those places more scientifically valuable to target if a survey's goal is to discover many resonant TNOs \citep[e.g.,][]{Lawler2013}.  These TNOs are also more dynamically coupled to Neptune and may have been swept up during Neptune's migration, thus giving extra constraints on Neptune migration simulations, both in orbital distributions \citep[e.g.][]{Gladman2012}, 
and in physical properties/formation location \citep{Pike2023}.

\subsubsection{Classical TNOs} \label{sec:classical}

The classical TNOs are most similar to the originally predicted dynamically cold, flat disk beyond Neptune \citep{Kuiper1951}.  Classical TNOs are those that are non-resonant and low-$e$ \citep[][uses $e<0.24$]{Gladman2008}.  It is currently unclear where the large-$a$ boundary should lay, originally it was posited that none would be found beyond the 2:1 resonance at $a=47.8$~AU, but the a few TNOs appear in the classical $e$ range just outside the 2:1 \citep[e.g.][]{Bannister2018}, perhaps implying that this edge is not primordial, but sculpted by planetary interactions - further discoveries are needed in the classical TNO region to understand exactly where it makes sense to put the outer edge of the classification.  There are several sub-classes within the classical TNOs, each giving their own clues to the past history of the Solar System \citep{Petit2011}.  The cold classical TNOs are one sub-class that is particularly useful for understanding Solar System history.

Cold classical TNOs also have low inclinations ($i\lesssim5$ degrees) and remain close to where they originally formed, and the fact that they never received violent dynamical kicks from the giant planets, like much of the rest of the TNO populations, means that they have a much higher fraction of binaries \citep{Fraser2017}.  This high rate of binarity in the primordial cold classical population is one of the important lines of evidence supporting the streaming instability mechanism (Section~\ref{sec:form}).

\subsubsection{Detached TNOs}

The detached TNOs are non-resonant and high-$e$ \citep[][uses $e>0.24$]{Gladman2008}.  These TNOs experience only very weak interactions with Neptune, though it has been demonstrated that these weak interactions can cause diffusion in $q$ and $a$ over long timescales, even out to $q\sim50$~AU \citep{Bannister2017}.  These TNOs were likely first emplaced by scattering off Neptune during the migration phase, and then ``fossilized'' on distant, fairly high-$q$ orbits as Neptune's orbit circularized.

\subsubsection{Sednoids}

There are now a handful of known TNOs now that have very high-$q$ ($q\gtrsim50-60$~AU), without large enough semimajor axes for Galactic tides to be the cause, and with no current resonant interactions with Neptune.  We propose these should be part of a new distant Sednoid TNO class, as discussed in \citet{Pike2017} and \citet{Lawler2019}.
There has recently been some discussion in the literature about so-called ``extreme TNOs,'' though different groups have used different $q$ and $a$ cuts to define this (typically relatively low $q$ and $a$ cuts, e.g., $q>37$~AU and $a>150$~AU).  These ``extreme TNOs'' have an apparent clustering in $\omega$ and/or $\omega+\Omega$ that has been used to argue for an undiscovered distant giant planet \citep[popularly referred to as ``Planet 9'';][]{Batygin2016}, though the conclusion from well-characterized TNO surveys like OSSOS and DES is that the apparent clustering is only an artifact of observational biases \citep{Kavelaars2020,Napier2021}.  We prefer a much higher cut in $q>60$~AU, as recommended by \citet{Huang2024}, to define the Sednoids.  There is a clear group of three very high-$q$, large-$a$ TNOs that are very different in orbital properties from all other known TNOs.
These 3 real Sednoids were extremely difficult to discover due to their always-large distances, implying a much larger population that exists beyond current detection limits \citep{Kavelaars2020}.  

While a few high-$q$ or near-high-$q$ TNOs can have their emplacement explained by past interactions with resonances \citep{Lawler2019}, Sednoids are actually the TNO population that we generally understand the least about.  However, there are a wide range of theories that could possibly explain these orbits.  Some possibilities include Sednoids being produced by close past stellar flybys \citep{Kenyon2004,Huang2024}, 
or self-gravitation in a massive planetesimal disk \citep{Madigan2018}, or even a rogue planet \citep{Huang2022}.  The only way we will discover what story of the Solar System's past history the Sednoids are trying to tell us, is by discovering more of them, which requires careful surveys that have very faint brightness limits, such the planned limits of the in-progress DEEP and CLASSY Surveys.

\subsection{TNO Size-Frequency Distributions} \label{sec:sizes}

TNOs come in a range of measured sizes, with Pluto being the largest currently, with a diameter of over 2,000~km.  The smallest TNO with a directly measured size is Arrokoth at 30~km, or Pluto's smallest moon Kerebos at 19~km. These were measured very accurately as the \textit{New Horizons} Mission probe flew past, but we (very unfortunately) can't send spacecraft to all TNOs.  We expect to find many more small objects than large objects in the Kuiper Belt, simply because if you break a large object into smaller pieces, there will be many more small pieces than large pieces.  But the exact distribution of sizes is not easy to predict, and depends on a number of factors.

Because very few TNOs are able to be resolved, the sizes must be inferred indirectly.  Clever measurements of stellar occultations can be used to measure the size and shape of a few TNOs, but this requires extreme orbital accuracy and a chance alignment between a decently bright star and a TNO, that is accessible from a well-populated part of Earth (more detail on thie technique in Section~\ref{sec:surf_props}).
Most TNOs have their sizes measured using albedo, distance, and apparent magnitude, and most albedo measurements have large uncertainties, so assumptions are made. But overall, the size distribution can still be measured, as long the assumptions are made carefully and consistently.  Measuring the size distribution accurately also requires debiasing, as for orbital distributions, because surveys will always be better at detecting the largest TNOs.  But again, if your survey magnitude limits are known, you can account for this bias.  Solar system absolute magnitude $H$ is often measured as a proxy for size, since it is a way to remove distance from the measurement - $H$ magnitude is the magnitude that a TNO would have if viewed by an observer sitting on the Sun (not recommended), observing the TNO sitting at 1~AU distance.  Size distributions are often parameterized in terms of $H$ magnitudes, and take the format 

\begin{align}
N(<H) \propto 10^{\alpha H}
\end{align} \label{eq:res}

where $N(<H)$ means the cumulative number of TNOs with a smaller $H$ magnitude than the limit given - because these are magnitudes, larger $H$ magnitudes correspond to smaller size TNOs.  
The largest TNOs have $H$ magnitudes around 0, while typical TNOs discovered in large TNO surveys (Table~\ref{tab:surveys}) are in the range of $H\simeq8-10$. Size distributions are well-measured for the largest TNOs, down to approximately 100~km in diameter \citep{Kavelaars2021}, and DEEP has now measured the size distribution down to even smaller sizes for some TNO populations \citep{Napier2024}.  The $\alpha$ exponent to the power-law size distribution is not constant for all sizes.  The most recent work suggests that $\alpha$ tapers off at smaller sizes, as a result of past collisions.
Measuring the size distribution of TNOs tells us both about formation - how big did planetesimals originally form (Section~\ref{sec:form}), and collisional evolution - how effectively have TNOs smashed together over the history of the Solar System.

\subsection{Surface Properties} \label{sec:surf_props}

The surface properties of TNOs depend on their composition and shape.
TNO surface reflectance is studied using multi-band photometry and spectroscopy.
Expanding photometry into thermal (infrared) bands, where TNOs are visible in emission instead of reflectance makes it possible to measure albedo and determine size instead of absolute magnitude.
The size of objects can also be measured using occultations, where the TNO passes in front of a star as viewed from the Earth. 
These size measurements are often paired with light-curve analysis, using photometry taken over long time baselines (hours to months) to determine the relative shape and rotation rate of objects.
These varied observing strategies have provided significant insight into the physical properties of TNOs.

An important physical property of TNOs is their shape, which can be round, extended, bi-lobed, heavily cratered, or irregular.
TNOs in general rotate with periods ranging from around five hours to more than 20 hours, with a range of amplitudes - some variability is obvious and some is extremely low amplitude \citep[e.g.][]{Benecchi2013,Thirouin2022} 
As an example, we discuss here the investigation of Arrokoth's shape prior to the \textit{New Horizons} Mission flyby, but these techniques can be applied generally to understand the shape and rotation of objects viewable only from Earth, and not with spacecraft.
The flyby images of Arrokoth clearly show it as a bi-lobed object, but this was not a complete surprise when the \textit{New Horizons} images were acquired - the research team already had hints of this shape from occultations!
The \textit{New Horizons} team used multiple methods to predict the shape and size of Arrokoth before encounter, including both light curve analysis \citep{Benecchi2019} and occultations \citep{Buie2020}.
The light curve analysis was done by acquiring many observations from the Hubble Space Telescope (HST) and measuring the object's brightness.
This revealed that the variability was quite small, $\le$0.15 magnitudes, which the team concluded meant that either Arrokoth was nearly spherical, or its rotation axis was nearly aligned with the line-of-sight from Earth \citep{Benecchi2019}.
The team also did observations of several occultation events, where Arrokoth passed in front of a star as viewed from certain places on Earth. 
Many observing stations were used to measure multiple lines or cords through Arrokoth as it passed in front of a star from slightly different perspectives, allowing recreation of the its shape by measuring the differing amounts of time the star was blocked by Arrokoth in each location.
These events have to be carefully predicted in advance, and observers sent to specific locations on Earth where the event will be visible.
One occultation of Arrokoth from 2017 July showed a clearly non-spherical shape, which is extremely well fit by a bi-lobed object or a close binary object \citep{Buie2020}.
The combination of measuring the light-curve and occultations provided an excellent prediction of the shape of Arrokoth before encounter, and these techniques are used broadly in TNO science.

\begin{figure}[t]
\centering
\includegraphics[width=.6\textwidth]{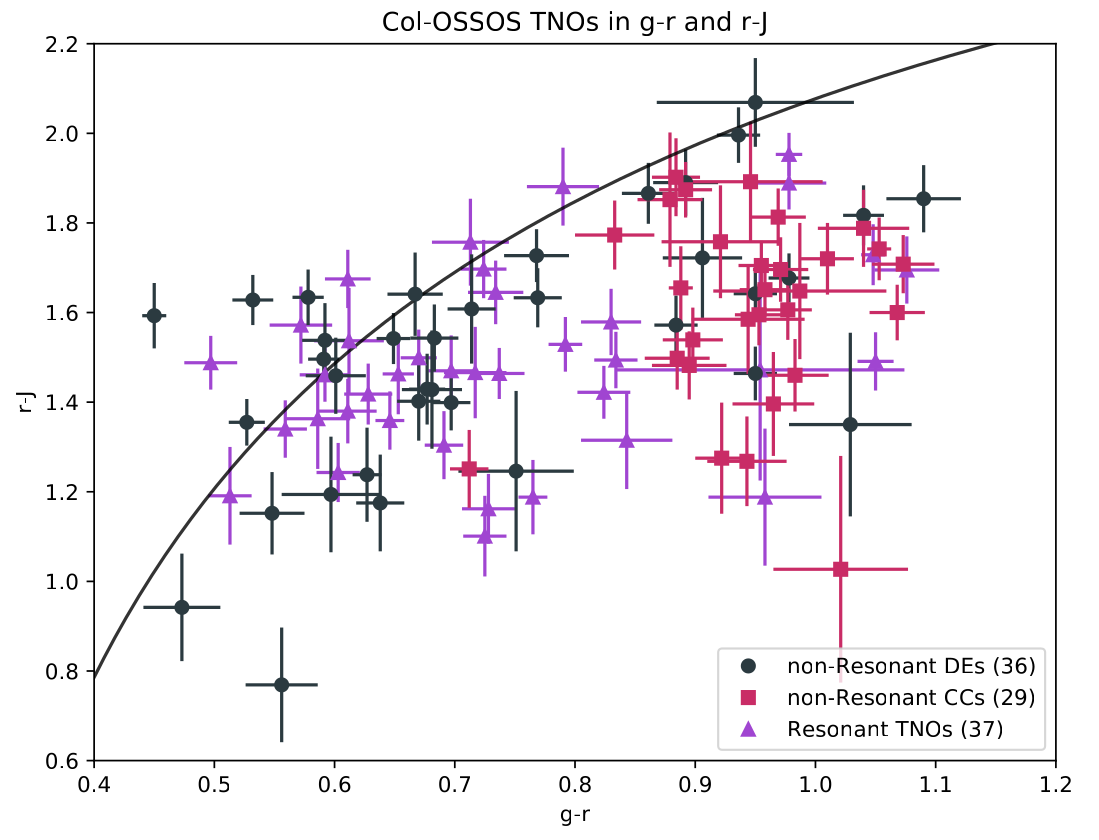}
\caption{The optical and near-infrared colors of TNOs, adapted from \citet{Pike2023}. The $r-J$ color (the difference between the brightness in optically red $r$-band and near-infrared $J$-band) is compared to the $g-r$ color (the difference between brightness in blue $g$-band and optical red $r$-band).  As these are magnitude differences, and larger means fainter, a larger number means that there is \textit{less} flux coming from the first band.  The cold classical objects (red squares, ``CCs'') are primarily optically red (larger $g-r$) and lower $r-J$.  The rest of the TNOs are divided into resonant objects, as described in Section \ref{sec:orbclass} and dynamically excited (``DE'') objects, which refers to all non-resonant non-cold classical TNOs.}
\label{fig:colors}
\end{figure}

Early work exploring TNO surfaces used multi-band photometry (measuring brightness in different colors wavelength bands) and found that TNO surfaces have a broad range of optical colors.
The bulk properties of smaller TNOs have been observed using multi-band photometry in order to classify their surfaces \citep[e.g.][]{Tegler2003,DalleOre2013,Peixinho2015,Fraser2023}.
Early studies noted that TNOs were all redder than solar, and their distribution had an apparent bimodality, with the two groups often referred to as `red' and `neutral' or `blue' \citep[e.g.][]{Tegler2003,Peixinho2003}. 
It was quickly noted that the overwhelming majority of objects with cold classical orbits (low-$e$ and low-$i$) were preferentially optically red \citep{Tegler2003}.
Unfortunately, these overlapped significantly with optically red objects in other dynamical classes.
Recent work has found that the combination of optical and near-infrared colors can be used to more clearly separate the cold classical surface type, which is optically red and less reflective in the near-infrared as compared to the optical \citep{Pike2023,Fraser2023}.
There is still some overlap, as shown in Figure \ref{fig:colors}, but the inclusion of a near-infrared band allows for a much more clean division between the surface types, called BrightIR and FaintIR in \citet{Fraser2023}.
These surface types are used to infer likely formation locations of the objects, with the FaintIR expected to form near the current cold classical region and slightly sunward, and the BrightIR to have formed sunward of that population \citep{Pike2023}.
The classification of TNOs based on their photometric colors has provided a useful tool for understanding their likely formation locations.

An additional complexity to understanding TNO surfaces is the range of albedos, or fraction of light reflected by the objects, they exhibit \citep[e.g.][]{Lellouch2013}.
A major argument for the uniqueness of the surfaces of cold classicals (and objects that formed near them) is their higher albedos of approximately 14\% compared to the other dynamical classes which had typical albedos of 8.5\% \citep{Vilenius2014}.
Albedos are measured by acquiring thermal data from infrared telescopes like the \textit{Herschel Space Observatory} and the \textit{Spitzer Space Telescope}.
In these wavelengths, TNOs are viewed purely in emission and not reflectance, so with the additional of optical magnitudes and thermal modeling the albedo and size of the objects can be calculated \citep{Vilenius2014}.
A general understanding of how apparent and absolute magnitudes convert to physical size for different TNO populations is critical to understanding how TNOs formed, for comparison with simulations of object formation.

Historically, only larger TNOs could be examined with spectroscopy \citep[e.g.][]{Barucci2008}, but with the recent launch of JWST, the number TNOs which can be observed spectroscopically has increased dramatically.
Large TNOs, such as Pluto, Eris, Sedna, and other dwarf planets were found to have many different surface ices including methane \citep{Barucci2005,Brown2005}.
Other ices, including water \citep[e.g.][]{Trujillo2007}, ammonia \citep{Jewitt2004}, and nitrogen \citep[e.g.][]{Brown2002,Trujillo2007}
indicate that large TNOs are largely volatile-rich objects.
Spectroscopic observations of mid-sized TNOs from the ground showed occasional water-ice and a featureless continuum, typically with a red slope in the optical \citep[e.g.][]{Barkume2008}.
JWST has begun to revolutionize our understanding of TNO surfaces.
The DiScovering the Composition of the trans-Neptunian objects (DiSCo-TNOs) project is acquiring spectroscopy of a carefully selected sample of 59 TNOs and Centaurs \citep{Pinilla-Alonso2021}.
Preliminary results from this program show several groups of TNO surface classes which are differentiated by their behaviour in the infrared, which seems to correlate to optical colors as well \citep{Pinilla-Alonso2024, DePra2024}.
TNOs in general appear rich in CO$_2$ and CO \citep{DePra2024}, and more exciting insights into the composition of the bulk population of TNOs and how the composition constrains formation are on the horizon.

\section{The smallest particles: dust} \label{sec:dust}

With more than 5,000 known exoplanets as of 2024\footnote{\url{https://exoplanetarchive.ipac.caltech.edu/}}, we are just approaching the point where we can place our Solar System within the larger population of planetary systems \citep{Winn2015}. 
In exoplanetary systems, we can also measure small bodies like those in Kuiper Belt analogues, but only indirectly through the collisionally produced dust. This dust is sometimes directly resolvable but more often appears as an excess of infrared emission in the stellar spectrum. The dust around a star is quite low-mass, but the high surface area efficiently absorbs stellar radiation and re-radiates at the dust grains' blackbody equilibrium temperature, giving some additional information on orbital properties of the emitting dust \citep{Hughes2018}.  We can also measure the dust population in our Solar System resulting from collisions within the Kuiper Belt and sublimating comets, though it is 2-3 orders of magnitude fainter than we could measure in an exoplanetary system \citep{Poppe2019}.  However, just after the dramatic reshuffling that occurred early in the Solar System's history, the dust produced in the Kuiper Belt would have been much closer in brightness to measured debris disk systems \citep{Booth2009}.

Within our own Solar System, measuring dust from the Kuiper Belt has been very difficult from Earth-based telescopes, as this extremely diffuse, cold emission is drowned out by closer, hotter dust from the asteroid belt \citep{Moro-Martin2003}.  The \textit{New Horizons} Mission is now travelling through the Kuiper Belt\footnote{\url{https://pluto.jhuapl.edu/Mission/Where-is-New-Horizons.php}}, and its dust counter instrument has been continually measuring small dust grains that physically hit the small detector.  The dust that the instrument is detecting now as the probe approaches 60~AU (twice the distance of Neptune from the Sun) should be generated by collisions between interplanetary dust and TNOs, as well as by the very infrequent collisions between TNOs, and is expected to be dropping off quickly since the probe has long since passed through the densest part of the Kuiper Belt.  However, the dust counts remain much higher than expected \citep{Doner2024}, adding to hints from very deep TNO surveys that the Kuiper Belt may extend to much larger distances than previously thought \citep{Fraser2023NH}.

\section{Conclusion and Future Outlook} \label{sec:future}

Thanks to careful studies of both individual TNOs in detail, and various TNO subpopulations, we have a very good overall understanding of how TNOs were formed and emplaced onto their current orbits by the migration of Neptune in the early history of the Solar System \citep{Gladman2021}.  
The size distribution of TNOs tells us about planetesimal formation, and gives constraints on the streaming instability and pebble accretion. 
The color and composition of individual objects and populations tells us about formation location and past migration history. 
There is, of course, a great deal more that we can learn from upcoming surveys, and a few key outstanding questions that still need to be answered.

Our understanding of the implications of the JWST data for TNOs, discussed briefly in Section \ref{sec:surf_props}, is still in its early stages.
Additional absorption features are likely to be identified in the TNO spectra, and as data of more objects is acquired, the chances of unexpected discoveries increases.
Future work will also likely involve using detailed infrared spectra to determine a set of diagnostic JWST filters for multi-band photometry on JWST, which will open up the spectral classification system to a huge number of TNOs which are too faint for spectroscopy.
The JWST results will likely also inspire additional laboratory studies on the ices found in the distant Solar System, which will aid in interpretation of these results.

The Legacy Survey of Space and Time (LSST), operated on the Vera C. Rubin Observatory, will also provide an unprecedented number of object discoveries for the outer solar system.
Current predictions are for LSST to discover 40,000 TNOs, mostly in the first 2 years of survey operation, scheduled to begin in 2025 \citep{VeraC.RubinObservatoryLSSTSolarSystemScienceCollaboration2021}.
These discoveries will provide a wealth of objects for followup characterization using JWST and other telescopic resources, as well as precise orbits for the thousands of discoveries.
LSST may reveal previously unknown populations, such as additional distant belts, new dwarf planets, or more distant high-$q$ or extreme TNOs which do not interact with Neptune.
As long as satellite pollution does not significantly hamper ground-based astronomy \citep{Eggl2023}, these exciting discoveries will revolutionize our understanding of the outer Solar System over the next 10 years!

\begin{ack}[Acknowledgments]

SML acknowledges funding from the Natural Sciences and Engineering Research Council of Canada. REP acknowledges funding from NASA Solar System Observations grant 80NSSC21K0289 and NASA Emerging Worlds grant 80NSSC21K0376.  This research used the Canadian Advanced Network For Astronomy Research (CANFAR) operated in partnership by the  Canadian Astronomy Data Centre and The Digital Research Alliance of Canada with support from the National Research Council of Canada, the Canadian Space Agency, CANARIE, and the Canadian Foundation for Innovation.  We are grateful to have access to telescopes on remote mountains for science observing, and we acknowledge the history of colonialism behind many telescope sites used for TNO research. We support better, more respectful engagement with Indigenous peoples going forward, as well as learning and sharing Indigenous traditional astronomy knowledge\footnote{A great resource to learn more about Indigenous starlore and traditional knowledge of the sky: \url{https://nativeskywatchers.com/}}.

\end{ack}


\bibliographystyle{Harvard}
\begin{thebibliography*}{118}
\providecommand{\bibtype}[1]{}
\providecommand{\natexlab}[1]{#1}
{\catcode`\|=0\catcode`\#=12\catcode`\@=11\catcode`\\=12
|immediate|write|@auxout{\expandafter\ifx\csname natexlab\endcsname\relax\gdef\natexlab#1{#1}\fi}}
\renewcommand{\url}[1]{{\tt #1}}
\providecommand{\urlprefix}{URL }
\expandafter\ifx\csname urlstyle\endcsname\relax
  \providecommand{\doi}[1]{doi:\discretionary{}{}{}#1}\else
  \providecommand{\doi}{doi:\discretionary{}{}{}\begingroup \urlstyle{rm}\Url}\fi
\providecommand{\bibinfo}[2]{#2}
\providecommand{\eprint}[2][]{\url{#2}}

\bibtype{Inproceedings}%
\bibitem[{Alexandersen} et al.(2023)]{Alexandersen2023}
\bibinfo{author}{{Alexandersen} M}, \bibinfo{author}{{Lawler} S}, \bibinfo{author}{{Chen} YT}, \bibinfo{author}{{Pike} R}, \bibinfo{author}{{Semenchuck} C}, \bibinfo{author}{{Comte} M}, \bibinfo{author}{{Peltier} L}, \bibinfo{author}{{Kavelaars} J} and  \bibinfo{author}{{Collyer} C} (\bibinfo{year}{2023}), \bibinfo{month}{Oct.}, \bibinfo{title}{{LiDO; orbital classification from precision tracking}}, \bibinfo{booktitle}{AAS/Division for Planetary Sciences Meeting Abstracts}, \bibinfo{series}{AAS/Division for Planetary Sciences Meeting Abstracts}, \bibinfo{volume}{55}, pp. \bibinfo{pages}{209.08}.

\bibtype{Article}%
\bibitem[Anderson and Darling(1952)]{Anderson1952}
\bibinfo{author}{Anderson TW} and  \bibinfo{author}{Darling DA} (\bibinfo{year}{1952}).
\bibinfo{title}{{Asymptotic Theory of Certain "Goodness of Fit" Criteria Based on Stochastic Processes}}.
\bibinfo{journal}{{\em The Annals of Mathematical Statistics}} \bibinfo{volume}{23} (\bibinfo{number}{2}): \bibinfo{pages}{193 -- 212}. \bibinfo{doi}{\doi{10.1214/aoms/1177729437}}.
\bibinfo{url}{\url{https://doi.org/10.1214/aoms/1177729437}}.

\bibtype{Article}%
\bibitem[{Andrews}(2020)]{Andrews2020}
\bibinfo{author}{{Andrews} SM} (\bibinfo{year}{2020}), \bibinfo{month}{Aug.}
\bibinfo{title}{{Observations of Protoplanetary Disk Structures}}.
\bibinfo{journal}{{\em \araa}} \bibinfo{volume}{58}: \bibinfo{pages}{483--528}. \bibinfo{doi}{\doi{10.1146/annurev-astro-031220-010302}}.
\eprint{2001.05007}.

\bibtype{Article}%
\bibitem[{Bailey} and {Fabrycky}(2019)]{Bailey2019}
\bibinfo{author}{{Bailey} N} and  \bibinfo{author}{{Fabrycky} D} (\bibinfo{year}{2019}), \bibinfo{month}{Aug.}
\bibinfo{title}{{Stellar Flybys Interrupting Planet-Planet Scattering Generates Oort Planets}}.
\bibinfo{journal}{{\em \aj}} \bibinfo{volume}{158} (\bibinfo{number}{2}), \bibinfo{eid}{94}. \bibinfo{doi}{\doi{10.3847/1538-3881/ab2d2a}}.
\eprint{1905.07044}.

\bibtype{Article}%
\bibitem[{Bannister} et al.(2017)]{Bannister2017}
\bibinfo{author}{{Bannister} MT}, \bibinfo{author}{{Shankman} C}, \bibinfo{author}{{Volk} K}, \bibinfo{author}{{Chen} YT}, \bibinfo{author}{{Kaib} N}, \bibinfo{author}{{Gladman} BJ}, \bibinfo{author}{{Jakubik} M}, \bibinfo{author}{{Kavelaars} JJ}, \bibinfo{author}{{Fraser} WC}, \bibinfo{author}{{Schwamb} ME}, \bibinfo{author}{{Petit} JM}, \bibinfo{author}{{Wang} SY}, \bibinfo{author}{{Gwyn} SDJ}, \bibinfo{author}{{Alexandersen} M} and  \bibinfo{author}{{Pike} RE} (\bibinfo{year}{2017}), \bibinfo{month}{Jun.}
\bibinfo{title}{{OSSOS. V. Diffusion in the Orbit of a High-perihelion Distant Solar System Object}}.
\bibinfo{journal}{{\em \aj}} \bibinfo{volume}{153} (\bibinfo{number}{6}), \bibinfo{eid}{262}. \bibinfo{doi}{\doi{10.3847/1538-3881/aa6db5}}.
\eprint{1704.01952}.

\bibtype{Article}%
\bibitem[{Bannister} et al.(2018)]{Bannister2018}
\bibinfo{author}{{Bannister} MT}, \bibinfo{author}{{Gladman} BJ}, \bibinfo{author}{{Kavelaars} JJ}, \bibinfo{author}{{Petit} JM}, \bibinfo{author}{{Volk} K}, \bibinfo{author}{{Chen} YT}, \bibinfo{author}{{Alexandersen} M}, \bibinfo{author}{{Gwyn} SDJ}, \bibinfo{author}{{Schwamb} ME}, \bibinfo{author}{{Ashton} E}, \bibinfo{author}{{Benecchi} SD}, \bibinfo{author}{{Cabral} N}, \bibinfo{author}{{Dawson} RI}, \bibinfo{author}{{Delsanti} A}, \bibinfo{author}{{Fraser} WC}, \bibinfo{author}{{Granvik} M}, \bibinfo{author}{{Greenstreet} S}, \bibinfo{author}{{Guilbert-Lepoutre} A}, \bibinfo{author}{{Ip} WH}, \bibinfo{author}{{Jakubik} M}, \bibinfo{author}{{Jones} RL}, \bibinfo{author}{{Kaib} NA}, \bibinfo{author}{{Lacerda} P}, \bibinfo{author}{{Van Laerhoven} C}, \bibinfo{author}{{Lawler} S}, \bibinfo{author}{{Lehner} MJ}, \bibinfo{author}{{Lin} HW}, \bibinfo{author}{{Lykawka} PS}, \bibinfo{author}{{Marsset} M}, \bibinfo{author}{{Murray-Clay} R}, \bibinfo{author}{{Pike} RE}, \bibinfo{author}{{Rousselot} P},
  \bibinfo{author}{{Shankman} C}, \bibinfo{author}{{Thirouin} A}, \bibinfo{author}{{Vernazza} P} and  \bibinfo{author}{{Wang} SY} (\bibinfo{year}{2018}), \bibinfo{month}{May}.
\bibinfo{title}{{OSSOS. VII. 800+ Trans-Neptunian Objects{\textemdash}The Complete Data Release}}.
\bibinfo{journal}{{\em \apjs}} \bibinfo{volume}{236} (\bibinfo{number}{1}), \bibinfo{eid}{18}. \bibinfo{doi}{\doi{10.3847/1538-4365/aab77a}}.
\eprint{1805.11740}.

\bibtype{Article}%
\bibitem[{Barkume} et al.(2008)]{Barkume2008}
\bibinfo{author}{{Barkume} KM}, \bibinfo{author}{{Brown} ME} and  \bibinfo{author}{{Schaller} EL} (\bibinfo{year}{2008}), \bibinfo{month}{Jan.}
\bibinfo{title}{{Near-Infrared Spectra of Centaurs and Kuiper Belt Objects}}.
\bibinfo{journal}{{\em \aj}} \bibinfo{volume}{135} (\bibinfo{number}{1}): \bibinfo{pages}{55--67}. \bibinfo{doi}{\doi{10.1088/0004-6256/135/1/55}}.

\bibtype{Article}%
\bibitem[{Barucci} et al.(2005)]{Barucci2005}
\bibinfo{author}{{Barucci} MA}, \bibinfo{author}{{Cruikshank} DP}, \bibinfo{author}{{Dotto} E}, \bibinfo{author}{{Merlin} F}, \bibinfo{author}{{Poulet} F}, \bibinfo{author}{{Dalle Ore} C}, \bibinfo{author}{{Fornasier} S} and  \bibinfo{author}{{de Bergh} C} (\bibinfo{year}{2005}), \bibinfo{month}{Aug.}
\bibinfo{title}{{Is Sedna another Triton?}}
\bibinfo{journal}{{\em \aap}} \bibinfo{volume}{439} (\bibinfo{number}{2}): \bibinfo{pages}{L1--L4}. \bibinfo{doi}{\doi{10.1051/0004-6361:200500144}}.

\bibtype{incollection}%
\bibitem[{Barucci} et al.(2008)]{Barucci2008}
\bibinfo{author}{{Barucci} MA}, \bibinfo{author}{{Brown} ME}, \bibinfo{author}{{Emery} JP} and  \bibinfo{author}{{Merlin} F} (\bibinfo{year}{2008}), \bibinfo{title}{{Composition and Surface Properties of Transneptunian Objects and Centaurs}}, \bibinfo{editor}{{Barucci} MA}, \bibinfo{editor}{{Boehnhardt} H}, \bibinfo{editor}{{Cruikshank} DP}, \bibinfo{editor}{{Morbidelli} A} and  \bibinfo{editor}{{Dotson} R}, (Eds.), \bibinfo{booktitle}{The Solar System Beyond Neptune},  \bibinfo{pages}{143--160}.

\bibtype{Article}%
\bibitem[{Bassa} et al.(2022)]{Bassa2022}
\bibinfo{author}{{Bassa} CG}, \bibinfo{author}{{Hainaut} OR} and  \bibinfo{author}{{Galad{\'\i}-Enr{\'\i}quez} D} (\bibinfo{year}{2022}), \bibinfo{month}{Jan.}
\bibinfo{title}{{Analytical simulations of the effect of satellite constellations on optical and near-infrared observations}}.
\bibinfo{journal}{{\em \aap}} \bibinfo{volume}{657}, \bibinfo{eid}{A75}. \bibinfo{doi}{\doi{10.1051/0004-6361/202142101}}.
\eprint{2108.12335}.

\bibtype{Article}%
\bibitem[{Batygin} and {Brown}(2016)]{Batygin2016}
\bibinfo{author}{{Batygin} K} and  \bibinfo{author}{{Brown} ME} (\bibinfo{year}{2016}), \bibinfo{month}{Feb.}
\bibinfo{title}{{Evidence for a Distant Giant Planet in the Solar System}}.
\bibinfo{journal}{{\em \aj}} \bibinfo{volume}{151} (\bibinfo{number}{2}), \bibinfo{eid}{22}. \bibinfo{doi}{\doi{10.3847/0004-6256/151/2/22}}.
\eprint{1601.05438}.

\bibtype{Article}%
\bibitem[{Batygin} et al.(2012)]{Batygin2012}
\bibinfo{author}{{Batygin} K}, \bibinfo{author}{{Brown} ME} and  \bibinfo{author}{{Betts} H} (\bibinfo{year}{2012}), \bibinfo{month}{Jan.}
\bibinfo{title}{{Instability-driven Dynamical Evolution Model of a Primordially Five-planet Outer Solar System}}.
\bibinfo{journal}{{\em \apjl}} \bibinfo{volume}{744} (\bibinfo{number}{1}), \bibinfo{eid}{L3}. \bibinfo{doi}{\doi{10.1088/2041-8205/744/1/L3}}.
\eprint{1111.3682}.

\bibtype{Article}%
\bibitem[{Benecchi} and {Sheppard}(2013)]{Benecchi2013}
\bibinfo{author}{{Benecchi} SD} and  \bibinfo{author}{{Sheppard} SS} (\bibinfo{year}{2013}), \bibinfo{month}{May}.
\bibinfo{title}{{Light Curves of 32 Large Transneptunian Objects}}.
\bibinfo{journal}{{\em \aj}} \bibinfo{volume}{145} (\bibinfo{number}{5}), \bibinfo{eid}{124}. \bibinfo{doi}{\doi{10.1088/0004-6256/145/5/124}}.
\eprint{1301.5791}.

\bibtype{Article}%
\bibitem[{Benecchi} et al.(2019)]{Benecchi2019}
\bibinfo{author}{{Benecchi} SD}, \bibinfo{author}{{Porter} SB}, \bibinfo{author}{{Buie} MW}, \bibinfo{author}{{Zangari} AM}, \bibinfo{author}{{Verbiscer} AJ}, \bibinfo{author}{{Noll} KS}, \bibinfo{author}{{Stern} SA}, \bibinfo{author}{{Spencer} JR} and  \bibinfo{author}{{Parker} AH} (\bibinfo{year}{2019}), \bibinfo{month}{Dec.}
\bibinfo{title}{{The HST lightcurve of (486958) 2014 MU$_{69}$}}.
\bibinfo{journal}{{\em \icarus}} \bibinfo{volume}{334}: \bibinfo{pages}{11--21}. \bibinfo{doi}{\doi{10.1016/j.icarus.2019.01.023}}.
\eprint{1812.04758}.

\bibtype{Article}%
\bibitem[{Bernardinelli} et al.(2022)]{Bernardinelli2022}
\bibinfo{author}{{Bernardinelli} PH}, \bibinfo{author}{{Bernstein} GM}, \bibinfo{author}{{Sako} M}, \bibinfo{author}{{Yanny} B}, \bibinfo{author}{{Aguena} M}, \bibinfo{author}{{Allam} S}, \bibinfo{author}{{Andrade-Oliveira} F}, \bibinfo{author}{{Bertin} E}, \bibinfo{author}{{Brooks} D}, \bibinfo{author}{{Buckley-Geer} E}, \bibinfo{author}{{Burke} DL}, \bibinfo{author}{{Carnero Rosell} A}, \bibinfo{author}{{Carrasco Kind} M}, \bibinfo{author}{{Carretero} J}, \bibinfo{author}{{Conselice} C}, \bibinfo{author}{{Costanzi} M}, \bibinfo{author}{{da Costa} LN}, \bibinfo{author}{{De Vicente} J}, \bibinfo{author}{{Desai} S}, \bibinfo{author}{{Diehl} HT}, \bibinfo{author}{{Dietrich} JP}, \bibinfo{author}{{Doel} P}, \bibinfo{author}{{Eckert} K}, \bibinfo{author}{{Everett} S}, \bibinfo{author}{{Ferrero} I}, \bibinfo{author}{{Flaugher} B}, \bibinfo{author}{{Fosalba} P}, \bibinfo{author}{{Frieman} J}, \bibinfo{author}{{Garc{\'\i}a-Bellido} J}, \bibinfo{author}{{Gerdes} DW}, \bibinfo{author}{{Gruen} D},
  \bibinfo{author}{{Gruendl} RA}, \bibinfo{author}{{Gschwend} J}, \bibinfo{author}{{Hinton} SR}, \bibinfo{author}{{Hollowood} DL}, \bibinfo{author}{{Honscheid} K}, \bibinfo{author}{{James} DJ}, \bibinfo{author}{{Kent} S}, \bibinfo{author}{{Kuehn} K}, \bibinfo{author}{{Kuropatkin} N}, \bibinfo{author}{{Lahav} O}, \bibinfo{author}{{Maia} MAG}, \bibinfo{author}{{March} M}, \bibinfo{author}{{Menanteau} F}, \bibinfo{author}{{Miquel} R}, \bibinfo{author}{{Morgan} R}, \bibinfo{author}{{Myles} J}, \bibinfo{author}{{Ogando} RLC}, \bibinfo{author}{{Palmese} A}, \bibinfo{author}{{Paz-Chinch{\'o}n} F}, \bibinfo{author}{{Pieres} A}, \bibinfo{author}{{Plazas Malag{\'o}n} AA}, \bibinfo{author}{{Romer} AK}, \bibinfo{author}{{Roodman} A}, \bibinfo{author}{{Sanchez} E}, \bibinfo{author}{{Scarpine} V}, \bibinfo{author}{{Schubnell} M}, \bibinfo{author}{{Serrano} S}, \bibinfo{author}{{Sevilla-Noarbe} I}, \bibinfo{author}{{Smith} M}, \bibinfo{author}{{Soares-Santos} M}, \bibinfo{author}{{Suchyta} E}, \bibinfo{author}{{Swanson}
  MEC}, \bibinfo{author}{{Tarle} G}, \bibinfo{author}{{To} C}, \bibinfo{author}{{Varga} TN} and  \bibinfo{author}{{Walker} AR} (\bibinfo{year}{2022}), \bibinfo{month}{Feb.}
\bibinfo{title}{{A Search of the Full Six Years of the Dark Energy Survey for Outer Solar System Objects}}.
\bibinfo{journal}{{\em \apjs}} \bibinfo{volume}{258} (\bibinfo{number}{2}), \bibinfo{eid}{41}. \bibinfo{doi}{\doi{10.3847/1538-4365/ac3914}}.
\eprint{2109.03758}.

\bibtype{Article}%
\bibitem[{Bernardinelli} et al.(2024)]{Bernardinelli2024}
\bibinfo{author}{{Bernardinelli} PH}, \bibinfo{author}{{Smotherman} H}, \bibinfo{author}{{Langford} Z}, \bibinfo{author}{{Portillo} SKN}, \bibinfo{author}{{Connolly} AJ}, \bibinfo{author}{{Kalmbach} JB}, \bibinfo{author}{{Stetzler} S}, \bibinfo{author}{{Juri{\'c}} M}, \bibinfo{author}{{Oldroyd} WJ}, \bibinfo{author}{{Lin} HW}, \bibinfo{author}{{Adams} FC}, \bibinfo{author}{{Chandler} CO}, \bibinfo{author}{{Fuentes} C}, \bibinfo{author}{{Gerdes} DW}, \bibinfo{author}{{Holman} MJ}, \bibinfo{author}{{Markwardt} L}, \bibinfo{author}{{McNeill} A}, \bibinfo{author}{{Mommert} M}, \bibinfo{author}{{Napier} KJ}, \bibinfo{author}{{Payne} MJ}, \bibinfo{author}{{Ragozzine} D}, \bibinfo{author}{{Rivkin} AS}, \bibinfo{author}{{Schlichting} H}, \bibinfo{author}{{Sheppard} SS}, \bibinfo{author}{{Strauss} R}, \bibinfo{author}{{Trilling} DE} and  \bibinfo{author}{{Trujillo} CA} (\bibinfo{year}{2024}), \bibinfo{month}{Mar.}
\bibinfo{title}{{The DECam Ecliptic Exploration Project (DEEP). III. Survey Characterization and Simulation Methods}}.
\bibinfo{journal}{{\em \aj}} \bibinfo{volume}{167} (\bibinfo{number}{3}), \bibinfo{eid}{134}. \bibinfo{doi}{\doi{10.3847/1538-3881/ad1527}}.

\bibtype{Article}%
\bibitem[{Bernstein} and {Khushalani}(2000)]{Bernstein2000}
\bibinfo{author}{{Bernstein} G} and  \bibinfo{author}{{Khushalani} B} (\bibinfo{year}{2000}), \bibinfo{month}{Dec.}
\bibinfo{title}{{Orbit Fitting and Uncertainties for Kuiper Belt Objects}}.
\bibinfo{journal}{{\em \aj}} \bibinfo{volume}{120} (\bibinfo{number}{6}): \bibinfo{pages}{3323--3332}. \bibinfo{doi}{\doi{10.1086/316868}}.
\eprint{astro-ph/0008348}.

\bibtype{Article}%
\bibitem[{Booth} et al.(2009)]{Booth2009}
\bibinfo{author}{{Booth} M}, \bibinfo{author}{{Wyatt} MC}, \bibinfo{author}{{Morbidelli} A}, \bibinfo{author}{{Moro-Mart{\'\i}n} A} and  \bibinfo{author}{{Levison} HF} (\bibinfo{year}{2009}), \bibinfo{month}{Oct.}
\bibinfo{title}{{The history of the Solar system's debris disc: observable properties of the Kuiper belt}}.
\bibinfo{journal}{{\em \mnras}} \bibinfo{volume}{399} (\bibinfo{number}{1}): \bibinfo{pages}{385--398}. \bibinfo{doi}{\doi{10.1111/j.1365-2966.2009.15286.x}}.
\eprint{0906.3755}.

\bibtype{Article}%
\bibitem[{Brown}(2001)]{Brown2001}
\bibinfo{author}{{Brown} ME} (\bibinfo{year}{2001}), \bibinfo{month}{May}.
\bibinfo{title}{{The Inclination Distribution of the Kuiper Belt}}.
\bibinfo{journal}{{\em \aj}} \bibinfo{volume}{121} (\bibinfo{number}{5}): \bibinfo{pages}{2804--2814}. \bibinfo{doi}{\doi{10.1086/320391}}.

\bibtype{Article}%
\bibitem[{Brown}(2002)]{Brown2002}
\bibinfo{author}{{Brown} ME} (\bibinfo{year}{2002}), \bibinfo{month}{Jan.}
\bibinfo{title}{{Pluto and Charon: Formation, Seasons, Composition}}.
\bibinfo{journal}{{\em Annual Review of Earth and Planetary Sciences}} \bibinfo{volume}{30}: \bibinfo{pages}{307--345}. \bibinfo{doi}{\doi{10.1146/annurev.earth.30.090401.095213}}.

\bibtype{Article}%
\bibitem[{Brown} and {Schaller}(2007)]{Brown2007}
\bibinfo{author}{{Brown} ME} and  \bibinfo{author}{{Schaller} EL} (\bibinfo{year}{2007}), \bibinfo{month}{Jun.}
\bibinfo{title}{{The Mass of Dwarf Planet Eris}}.
\bibinfo{journal}{{\em Science}} \bibinfo{volume}{316} (\bibinfo{number}{5831}): \bibinfo{pages}{1585}. \bibinfo{doi}{\doi{10.1126/science.1139415}}.

\bibtype{Article}%
\bibitem[{Brown} et al.(2004)]{Brown2004}
\bibinfo{author}{{Brown} ME}, \bibinfo{author}{{Trujillo} C} and  \bibinfo{author}{{Rabinowitz} D} (\bibinfo{year}{2004}), \bibinfo{month}{Dec.}
\bibinfo{title}{{Discovery of a Candidate Inner Oort Cloud Planetoid}}.
\bibinfo{journal}{{\em \apj}} \bibinfo{volume}{617} (\bibinfo{number}{1}): \bibinfo{pages}{645--649}. \bibinfo{doi}{\doi{10.1086/422095}}.
\eprint{astro-ph/0404456}.

\bibtype{Article}%
\bibitem[{Brown} et al.(2005)]{Brown2005}
\bibinfo{author}{{Brown} ME}, \bibinfo{author}{{Trujillo} CA} and  \bibinfo{author}{{Rabinowitz} DL} (\bibinfo{year}{2005}), \bibinfo{month}{Dec.}
\bibinfo{title}{{Discovery of a Planetary-sized Object in the Scattered Kuiper Belt}}.
\bibinfo{journal}{{\em \apjl}} \bibinfo{volume}{635} (\bibinfo{number}{1}): \bibinfo{pages}{L97--L100}. \bibinfo{doi}{\doi{10.1086/499336}}.
\eprint{astro-ph/0508633}.

\bibtype{Article}%
\bibitem[{Buchanan} et al.(2022)]{Buchanan2022}
\bibinfo{author}{{Buchanan} LE}, \bibinfo{author}{{Schwamb} ME}, \bibinfo{author}{{Fraser} WC}, \bibinfo{author}{{Bannister} MT}, \bibinfo{author}{{Marsset} M}, \bibinfo{author}{{Pike} RE}, \bibinfo{author}{{Nesvorn{\'y}} D}, \bibinfo{author}{{Kavelaars} JJ}, \bibinfo{author}{{Benecchi} SD}, \bibinfo{author}{{Lehner} MJ}, \bibinfo{author}{{Wang} SY}, \bibinfo{author}{{Peixinho} N}, \bibinfo{author}{{Volk} K}, \bibinfo{author}{{Alexandersen} M}, \bibinfo{author}{{Chen} YT}, \bibinfo{author}{{Gladman} B}, \bibinfo{author}{{Gwyn} S} and  \bibinfo{author}{{Petit} JM} (\bibinfo{year}{2022}), \bibinfo{month}{Jan.}
\bibinfo{title}{{Col-OSSOS: Probing Ice Line/Color Transitions within the Kuiper Belt's Progenitor Populations}}.
\bibinfo{journal}{{\em \psj}} \bibinfo{volume}{3} (\bibinfo{number}{1}), \bibinfo{eid}{9}. \bibinfo{doi}{\doi{10.3847/PSJ/ac42c9}}.
\eprint{2112.06754}.

\bibtype{Article}%
\bibitem[{Buie} et al.(2020)]{Buie2020}
\bibinfo{author}{{Buie} MW}, \bibinfo{author}{{Porter} SB}, \bibinfo{author}{{Tamblyn} P}, \bibinfo{author}{{Terrell} D}, \bibinfo{author}{{Parker} AH}, \bibinfo{author}{{Baratoux} D}, \bibinfo{author}{{Kaire} M}, \bibinfo{author}{{Leiva} R}, \bibinfo{author}{{Verbiscer} AJ}, \bibinfo{author}{{Zangari} AM}, \bibinfo{author}{{Colas} F}, \bibinfo{author}{{Diop} BD}, \bibinfo{author}{{Samaniego} JI}, \bibinfo{author}{{Wasserman} LH}, \bibinfo{author}{{Benecchi} SD}, \bibinfo{author}{{Caspi} A}, \bibinfo{author}{{Gwyn} S}, \bibinfo{author}{{Kavelaars} JJ}, \bibinfo{author}{{Ocampo Ur{\'\i}a} AC}, \bibinfo{author}{{Rabassa} J}, \bibinfo{author}{{Skrutskie} MF}, \bibinfo{author}{{Soto} A}, \bibinfo{author}{{Tanga} P}, \bibinfo{author}{{Young} EF}, \bibinfo{author}{{Stern} SA}, \bibinfo{author}{{Andersen} BC}, \bibinfo{author}{{Arango P{\'e}rez} ME}, \bibinfo{author}{{Arredondo} A}, \bibinfo{author}{{Artola} RA}, \bibinfo{author}{{B{\^a}} A}, \bibinfo{author}{{Ballet} R}, \bibinfo{author}{{Blank} T},
  \bibinfo{author}{{Bop} CT}, \bibinfo{author}{{Bosh} AS}, \bibinfo{author}{{Camino L{\'o}pez} MA}, \bibinfo{author}{{Carter} CM}, \bibinfo{author}{{Castro-Chac{\'o}n} JH}, \bibinfo{author}{{Caycedo Desprez} A}, \bibinfo{author}{{Caycedo Guerra} N}, \bibinfo{author}{{Conard} SJ}, \bibinfo{author}{{Dauvergne} JL}, \bibinfo{author}{{Dean} B}, \bibinfo{author}{{Dean} M}, \bibinfo{author}{{Desmars} J}, \bibinfo{author}{{Dieng} AL}, \bibinfo{author}{{Bousso Dieng} MD}, \bibinfo{author}{{Diouf} O}, \bibinfo{author}{{Dorego} GS}, \bibinfo{author}{{Dunham} DW}, \bibinfo{author}{{Dunham} J}, \bibinfo{author}{{Durantini Luca} HA}, \bibinfo{author}{{Edwards} P}, \bibinfo{author}{{Erasmus} N}, \bibinfo{author}{{Faye} G}, \bibinfo{author}{{Faye} M}, \bibinfo{author}{{Ferrario} LE}, \bibinfo{author}{{Ferrell} CL}, \bibinfo{author}{{Finley} TJ}, \bibinfo{author}{{Fraser} WC}, \bibinfo{author}{{Friedli} AJ}, \bibinfo{author}{{Galvez Serna} J}, \bibinfo{author}{{Garcia-Migani} EA}, \bibinfo{author}{{Genade} A},
  \bibinfo{author}{{Getrost} K}, \bibinfo{author}{{Gil-Hutton} RA}, \bibinfo{author}{{Gimeno} GN}, \bibinfo{author}{{Golub} EJ}, \bibinfo{author}{{Gonz{\'a}lez Murillo} GF}, \bibinfo{author}{{Grusin} MD}, \bibinfo{author}{{Gurovich} S}, \bibinfo{author}{{Hanna} WH}, \bibinfo{author}{{Henn} SM}, \bibinfo{author}{{Hinton} PC}, \bibinfo{author}{{Hughes} PJ}, \bibinfo{author}{{Josephs} John~David J}, \bibinfo{author}{{Joya} R}, \bibinfo{author}{{Kammer} JA}, \bibinfo{author}{{Keeney} BA}, \bibinfo{author}{{Keller} JM}, \bibinfo{author}{{Kramer} EA}, \bibinfo{author}{{Levine} SE}, \bibinfo{author}{{Lisse} CM}, \bibinfo{author}{{Lovell} AJ}, \bibinfo{author}{{Mackie} JA}, \bibinfo{author}{{Makarchuk} S}, \bibinfo{author}{{Manzano} LE}, \bibinfo{author}{{Mbaye} SS}, \bibinfo{author}{{Mbaye} M}, \bibinfo{author}{{Melia} RR}, \bibinfo{author}{{Moreno} F}, \bibinfo{author}{{Moss} SK}, \bibinfo{author}{{Ndaiye} D}, \bibinfo{author}{{Ndiaye} M}, \bibinfo{author}{{Nelson} MJ}, \bibinfo{author}{{Olkin} CB},
  \bibinfo{author}{{Olsen} AM}, \bibinfo{author}{{Ospina Moreno} VJ}, \bibinfo{author}{{Pasachoff} JM}, \bibinfo{author}{{Pereyra} MB}, \bibinfo{author}{{Person} MJ}, \bibinfo{author}{{Pinz{\'o}n} G}, \bibinfo{author}{{Pulver} EA}, \bibinfo{author}{{Quintero} EA}, \bibinfo{author}{{Regester} JR}, \bibinfo{author}{{Resnick} AC}, \bibinfo{author}{{Reyes-Ruiz} M}, \bibinfo{author}{{Rolfsmeier} AD}, \bibinfo{author}{{Ruhland} TR}, \bibinfo{author}{{Salmon} J}, \bibinfo{author}{{Santos-Sanz} P}, \bibinfo{author}{{Santucho} MA}, \bibinfo{author}{{Sep{\'u}lveda Ni{\~n}o} DK}, \bibinfo{author}{{Sickafoose} AA}, \bibinfo{author}{{Silva} JS}, \bibinfo{author}{{Singer} KN}, \bibinfo{author}{{Skipper} JN}, \bibinfo{author}{{Slivan} SM}, \bibinfo{author}{{Smith} RJC}, \bibinfo{author}{{Spagnotto} JC}, \bibinfo{author}{{Stephens} AW}, \bibinfo{author}{{Strabala} SD}, \bibinfo{author}{{Tamayo} FJ}, \bibinfo{author}{{Throop} HB}, \bibinfo{author}{{Torres Ca{\~n}as} AD}, \bibinfo{author}{{Toure} L}, \bibinfo{author}{{Traore}
  A}, \bibinfo{author}{{Tsang} CCC}, \bibinfo{author}{{Turner} JD}, \bibinfo{author}{{Vanegas} S}, \bibinfo{author}{{Venable} R}, \bibinfo{author}{{Wilson} JC}, \bibinfo{author}{{Zuluaga} CA} and  \bibinfo{author}{{Zuluaga} JI} (\bibinfo{year}{2020}), \bibinfo{month}{Apr.}
\bibinfo{title}{{Size and Shape Constraints of (486958) Arrokoth from Stellar Occultations}}.
\bibinfo{journal}{{\em \aj}} \bibinfo{volume}{159} (\bibinfo{number}{4}), \bibinfo{eid}{130}. \bibinfo{doi}{\doi{10.3847/1538-3881/ab6ced}}.
\eprint{2001.00125}.

\bibtype{Article}%
\bibitem[{Crompvoets} et al.(2022)]{Crompvoets2022}
\bibinfo{author}{{Crompvoets} BL}, \bibinfo{author}{{Lawler} SM}, \bibinfo{author}{{Volk} K}, \bibinfo{author}{{Chen} YT}, \bibinfo{author}{{Gladman} B}, \bibinfo{author}{{Peltier} L}, \bibinfo{author}{{Alexandersen} M}, \bibinfo{author}{{Bannister} MT}, \bibinfo{author}{{Gwyn} S}, \bibinfo{author}{{Kavelaars} JJ} and  \bibinfo{author}{{Petit} JM} (\bibinfo{year}{2022}), \bibinfo{month}{May}.
\bibinfo{title}{{OSSOS XXV: Large Populations and Scattering-Sticking in the Distant Trans-Neptunian Resonances}}.
\bibinfo{journal}{{\em \psj}} \bibinfo{volume}{3} (\bibinfo{number}{5}), \bibinfo{eid}{113}. \bibinfo{doi}{\doi{10.3847/PSJ/ac67e0}}.
\eprint{2204.09139}.

\bibtype{Article}%
\bibitem[{Dalle Ore} et al.(2013)]{DalleOre2013}
\bibinfo{author}{{Dalle Ore} CM}, \bibinfo{author}{{Dalle Ore} LV}, \bibinfo{author}{{Roush} TL}, \bibinfo{author}{{Cruikshank} DP}, \bibinfo{author}{{Emery} JP}, \bibinfo{author}{{Pinilla-Alonso} N} and  \bibinfo{author}{{Marzo} GA} (\bibinfo{year}{2013}), \bibinfo{month}{Jan.}
\bibinfo{title}{{A compositional interpretation of trans-neptunian objects taxonomies}}.
\bibinfo{journal}{{\em \icarus}} \bibinfo{volume}{222} (\bibinfo{number}{1}): \bibinfo{pages}{307--322}. \bibinfo{doi}{\doi{10.1016/j.icarus.2012.11.015}}.

\bibtype{Article}%
\bibitem[{De Pr{\'a}} et al.(2024)]{DePra2024}
\bibinfo{author}{{De Pr{\'a}} MN}, \bibinfo{author}{{H{\'e}nault} E}, \bibinfo{author}{{Pinilla-Alonso} N}, \bibinfo{author}{{Holler} BJ}, \bibinfo{author}{{Brunetto} R}, \bibinfo{author}{{Stansberry} JA}, \bibinfo{author}{{de Souza Feliciano} AC}, \bibinfo{author}{{Carvano} JM}, \bibinfo{author}{{Harvison} B}, \bibinfo{author}{{Licandro} J}, \bibinfo{author}{{M{\"u}ller} TG}, \bibinfo{author}{{Peixinho} N}, \bibinfo{author}{{Lorenzi} V}, \bibinfo{author}{{Guilbert-Lepoutre} A}, \bibinfo{author}{{Bannister} MT}, \bibinfo{author}{{Pendleton} YJ}, \bibinfo{author}{{Cruikshank} DP}, \bibinfo{author}{{Schambeau} CA}, \bibinfo{author}{{McClure} L} and  \bibinfo{author}{{Emery} JP} (\bibinfo{year}{2024}), \bibinfo{month}{May}.
\bibinfo{title}{{Widespread CO$_{2}$ and CO ices in the trans-Neptunian population revealed by JWST/DiSCo-TNOs}}.
\bibinfo{journal}{{\em Nature Astronomy}} \bibinfo{doi}{\doi{10.1038/s41550-024-02276-x}}.

\bibtype{Article}%
\bibitem[{Doner} et al.(2024)]{Doner2024}
\bibinfo{author}{{Doner} A}, \bibinfo{author}{{Hor{\'a}nyi} M}, \bibinfo{author}{{Bagenal} F}, \bibinfo{author}{{Brandt} P}, \bibinfo{author}{{Grundy} W}, \bibinfo{author}{{Lisse} C}, \bibinfo{author}{{Parker} J}, \bibinfo{author}{{Poppe} AR}, \bibinfo{author}{{Singer} KN}, \bibinfo{author}{{Stern} SA} and  \bibinfo{author}{{Verbiscer} A} (\bibinfo{year}{2024}), \bibinfo{month}{Feb.}
\bibinfo{title}{{New Horizons Venetia Burney Student Dust Counter Observes Higher than Expected Fluxes Approaching 60 au}}.
\bibinfo{journal}{{\em \apjl}} \bibinfo{volume}{961} (\bibinfo{number}{2}), \bibinfo{eid}{L38}. \bibinfo{doi}{\doi{10.3847/2041-8213/ad18b0}}.
\eprint{2401.01230}.

\bibtype{Inproceedings}%
\bibitem[{Dones} et al.(1996)]{Dones1996}
\bibinfo{author}{{Dones} L}, \bibinfo{author}{{Levison} HF} and  \bibinfo{author}{{Duncan} M} (\bibinfo{year}{1996}), \bibinfo{month}{Jan.}, \bibinfo{title}{{On the Dynamical Lifetimes of Planet--Crossing Objects}}, \bibinfo{editor}{{Rettig} T} and  \bibinfo{editor}{{Hahn} JM}, (Eds.), \bibinfo{booktitle}{Completing the Inventory of the Solar System}, \bibinfo{series}{Astronomical Society of the Pacific Conference Series}, \bibinfo{volume}{107},  \bibinfo{pages}{233--244}.

\bibtype{incollection}%
\bibitem[{Dones} et al.(2004)]{Dones2004}
\bibinfo{author}{{Dones} L}, \bibinfo{author}{{Weissman} PR}, \bibinfo{author}{{Levison} HF} and  \bibinfo{author}{{Duncan} MJ} (\bibinfo{year}{2004}), \bibinfo{title}{{Oort cloud formation and dynamics}}, \bibinfo{editor}{{Festou} MC}, \bibinfo{editor}{{Keller} HU} and  \bibinfo{editor}{{Weaver} HA}, (Eds.), \bibinfo{booktitle}{Comets II}, pp. \bibinfo{pages}{153}.

\bibtype{Article}%
\bibitem[{Duncan} and {Levison}(1997)]{Duncan1997}
\bibinfo{author}{{Duncan} MJ} and  \bibinfo{author}{{Levison} HF} (\bibinfo{year}{1997}), \bibinfo{month}{Jun.}
\bibinfo{title}{{A scattered comet disk and the origin of Jupiter family comets}}.
\bibinfo{journal}{{\em Science}} \bibinfo{volume}{276}: \bibinfo{pages}{1670--1672}. \bibinfo{doi}{\doi{10.1126/science.276.5319.1670}}.

\bibtype{Article}%
\bibitem[{Edgeworth}(1949)]{Edgeworth1949}
\bibinfo{author}{{Edgeworth} KE} (\bibinfo{year}{1949}), \bibinfo{month}{Jan.}
\bibinfo{title}{{The origin and evolution of the Solar System}}.
\bibinfo{journal}{{\em \mnras}} \bibinfo{volume}{109}: \bibinfo{pages}{600--609}. \bibinfo{doi}{\doi{10.1093/mnras/109.5.600}}.

\bibtype{Inproceedings}%
\bibitem[{Eggl} et al.(2023)]{Eggl2023}
\bibinfo{author}{{Eggl} S}, \bibinfo{author}{{Srivastava} S} and  \bibinfo{author}{{Cornwall} S} (\bibinfo{year}{2023}), \bibinfo{month}{Aug.}, \bibinfo{title}{{The Impact of Satellite Constellations on Solar System Science with LSST}}, \bibinfo{booktitle}{LPI Contributions}, \bibinfo{series}{LPI Contributions}, \bibinfo{volume}{2851}, pp. \bibinfo{pages}{2575}.

\bibtype{Article}%
\bibitem[{Fraser}(2024)]{Fraser2024}
\bibinfo{author}{{Fraser} WC} (\bibinfo{year}{2024}), \bibinfo{month}{May}.
\bibinfo{title}{{Detecting Moving Objects With Machine Learning}}.
\bibinfo{journal}{{\em arXiv e-prints}} , \bibinfo{eid}{arXiv:2405.06148}\bibinfo{doi}{\doi{10.48550/arXiv.2405.06148}}.
\eprint{2405.06148}.

\bibtype{Article}%
\bibitem[{Fraser} and {Brown}(2012)]{Fraser2012}
\bibinfo{author}{{Fraser} WC} and  \bibinfo{author}{{Brown} ME} (\bibinfo{year}{2012}), \bibinfo{month}{Apr.}
\bibinfo{title}{{The Hubble Wide Field Camera 3 Test of Surfaces in the Outer Solar System: The Compositional Classes of the Kuiper Belt}}.
\bibinfo{journal}{{\em \apj}} \bibinfo{volume}{749} (\bibinfo{number}{1}), \bibinfo{eid}{33}. \bibinfo{doi}{\doi{10.1088/0004-637X/749/1/33}}.
\eprint{1202.0827}.

\bibtype{Article}%
\bibitem[{Fraser} et al.(2017)]{Fraser2017}
\bibinfo{author}{{Fraser} WC}, \bibinfo{author}{{Bannister} MT}, \bibinfo{author}{{Pike} RE}, \bibinfo{author}{{Marsset} M}, \bibinfo{author}{{Schwamb} ME}, \bibinfo{author}{{Kavelaars} JJ}, \bibinfo{author}{{Lacerda} P}, \bibinfo{author}{{Nesvorn{\'y}} D}, \bibinfo{author}{{Volk} K}, \bibinfo{author}{{Delsanti} A}, \bibinfo{author}{{Benecchi} S}, \bibinfo{author}{{Lehner} MJ}, \bibinfo{author}{{Noll} K}, \bibinfo{author}{{Gladman} B}, \bibinfo{author}{{Petit} JM}, \bibinfo{author}{{Gwyn} S}, \bibinfo{author}{{Chen} YT}, \bibinfo{author}{{Wang} SY}, \bibinfo{author}{{Alexandersen} M}, \bibinfo{author}{{Burdullis} T}, \bibinfo{author}{{Sheppard} S} and  \bibinfo{author}{{Trujillo} C} (\bibinfo{year}{2017}), \bibinfo{month}{Apr.}
\bibinfo{title}{{All planetesimals born near the Kuiper belt formed as binaries}}.
\bibinfo{journal}{{\em Nature Astronomy}} \bibinfo{volume}{1}, \bibinfo{eid}{0088}. \bibinfo{doi}{\doi{10.1038/s41550-017-0088}}.
\eprint{1705.00683}.

\bibtype{Inproceedings}%
\bibitem[{Fraser} et al.(2023{\natexlab{a}})]{Fraser2023CLASSY}
\bibinfo{author}{{Fraser} WC}, \bibinfo{author}{{Lawler} S}, \bibinfo{author}{{Pike} RE}, \bibinfo{author}{{Kavelaars} J}, \bibinfo{author}{{Ashton} E}, \bibinfo{author}{{Gwyn} S}, \bibinfo{author}{{Chen} YT}, \bibinfo{author}{{Huang} Y}, \bibinfo{author}{{Gladman} B}, \bibinfo{author}{{Petit} JM}, \bibinfo{author}{{Semenchuck} C}, \bibinfo{author}{{Peltier} L}, \bibinfo{author}{{Alexandersen} M}, \bibinfo{author}{{Noyelles} B}, \bibinfo{author}{{Hestoffer} D}, \bibinfo{author}{{Chang} CK}, \bibinfo{author}{{Connolly} A}, \bibinfo{author}{{Kalmbach} JB}, \bibinfo{author}{{Wang} SY}, \bibinfo{author}{{Eduardo} M}, \bibinfo{author}{{Juric} M}, \bibinfo{author}{{Van Laerhoven} C}, \bibinfo{author}{{Bannister} M}, \bibinfo{author}{{Cowan} P}, \bibinfo{author}{{Tan} N} and  \bibinfo{author}{{Volk} K} (\bibinfo{year}{2023}{\natexlab{a}}), \bibinfo{month}{Aug.}, \bibinfo{title}{{The Classical and Large {\textemdash} A Solar System}}, \bibinfo{booktitle}{LPI Contributions}, \bibinfo{series}{LPI Contributions},
  \bibinfo{volume}{2851}, pp. \bibinfo{pages}{2346}.

\bibtype{Article}%
\bibitem[{Fraser} et al.(2023{\natexlab{b}})]{Fraser2023}
\bibinfo{author}{{Fraser} WC}, \bibinfo{author}{{Pike} RE}, \bibinfo{author}{{Marsset} M}, \bibinfo{author}{{Schwamb} ME}, \bibinfo{author}{{Bannister} MT}, \bibinfo{author}{{Buchanan} L}, \bibinfo{author}{{Kavelaars} JJ}, \bibinfo{author}{{Benecchi} SD}, \bibinfo{author}{{Tan} NJ}, \bibinfo{author}{{Peixinho} N}, \bibinfo{author}{{Gwyn} SDJ}, \bibinfo{author}{{Alexandersen} M}, \bibinfo{author}{{Chen} YT}, \bibinfo{author}{{Gladman} B} and  \bibinfo{author}{{Volk} K} (\bibinfo{year}{2023}{\natexlab{b}}), \bibinfo{month}{May}.
\bibinfo{title}{{Col-OSSOS: The Two Types of Kuiper Belt Surfaces}}.
\bibinfo{journal}{{\em \psj}} \bibinfo{volume}{4} (\bibinfo{number}{5}), \bibinfo{eid}{80}. \bibinfo{doi}{\doi{10.3847/PSJ/acc844}}.
\eprint{2206.04068}.

\bibtype{Inproceedings}%
\bibitem[{Fraser} et al.(2023{\natexlab{c}})]{Fraser2023NH}
\bibinfo{author}{{Fraser} WC}, \bibinfo{author}{{Porter} SB}, \bibinfo{author}{{Lin} HW}, \bibinfo{author}{{Napier} K}, \bibinfo{author}{{Spencer} RJ}, \bibinfo{author}{{Kavelaars} J}, \bibinfo{author}{{Verbiscer} AJ}, \bibinfo{author}{{Yoshida} F}, \bibinfo{author}{{Terai} T}, \bibinfo{author}{{Ito} T}, \bibinfo{author}{{Gerdes} D}, \bibinfo{author}{{Benecchi} SD}, \bibinfo{author}{{Stern} SA}, \bibinfo{author}{{Gwyn} S}, \bibinfo{author}{{Buie} MW}, \bibinfo{author}{{Peltier} L}, \bibinfo{author}{{Singer} KN}, \bibinfo{author}{{Brandy} PC}, \bibinfo{author}{{New Horizons Lorri Team}} and  \bibinfo{author}{{New Horizons Ggi Science Team}} (\bibinfo{year}{2023}{\natexlab{c}}), \bibinfo{month}{Mar.}, \bibinfo{title}{{Approaches to Detecting Kuiper Belt Objects for NASA's New Horizons Extended Mission: Digging Into the Noise}}, \bibinfo{booktitle}{54th Lunar and Planetary Science Conference}, \bibinfo{series}{LPI Contributions}, \bibinfo{volume}{2806}, pp. \bibinfo{pages}{2361}.

\bibtype{Article}%
\bibitem[{Gaia Collaboration} et al.(2016)]{GaiaCollaboration2016}
\bibinfo{author}{{Gaia Collaboration}}, \bibinfo{author}{{Brown} AGA}, \bibinfo{author}{{Vallenari} A}, \bibinfo{author}{{Prusti} T}, \bibinfo{author}{{de Bruijne} JHJ}, \bibinfo{author}{{Mignard} F}, \bibinfo{author}{{Drimmel} R}, \bibinfo{author}{{Babusiaux} C}, \bibinfo{author}{{Bailer-Jones} CAL}, \bibinfo{author}{{Bastian} U} and  \bibinfo{author}{et~al.} (\bibinfo{year}{2016}), \bibinfo{month}{Nov.}
\bibinfo{title}{{Gaia Data Release 1. Summary of the astrometric, photometric, and survey properties}}.
\bibinfo{journal}{{\em \aap}} \bibinfo{volume}{595}, \bibinfo{eid}{A2}. \bibinfo{doi}{\doi{10.1051/0004-6361/201629512}}.
\eprint{1609.04172}.

\bibtype{Article}%
\bibitem[{Gladman} and {Kavelaars}(1997)]{gladman1997}
\bibinfo{author}{{Gladman} B} and  \bibinfo{author}{{Kavelaars} JJ} (\bibinfo{year}{1997}), \bibinfo{month}{Jan.}
\bibinfo{title}{{Kuiper Belt searches from the Palomar 5-m telescope.}}
\bibinfo{journal}{{\em \aap}} \bibinfo{volume}{317}: \bibinfo{pages}{L35--L38}. \bibinfo{doi}{\doi{10.48550/arXiv.astro-ph/9610150}}.
\eprint{astro-ph/9610150}.

\bibtype{Article}%
\bibitem[{Gladman} and {Volk}(2021)]{Gladman2021}
\bibinfo{author}{{Gladman} B} and  \bibinfo{author}{{Volk} K} (\bibinfo{year}{2021}), \bibinfo{month}{Sep.}
\bibinfo{title}{{Transneptunian Space}}.
\bibinfo{journal}{{\em \araa}} \bibinfo{volume}{59}: \bibinfo{pages}{203--246}. \bibinfo{doi}{\doi{10.1146/annurev-astro-120920-010005}}.

\bibtype{incollection}%
\bibitem[{Gladman} et al.(2008)]{Gladman2008}
\bibinfo{author}{{Gladman} B}, \bibinfo{author}{{Marsden} BG} and  \bibinfo{author}{{Vanlaerhoven} C} (\bibinfo{year}{2008}), \bibinfo{title}{{Nomenclature in the Outer Solar System}}, \bibinfo{editor}{{Barucci} MA}, \bibinfo{editor}{{Boehnhardt} H}, \bibinfo{editor}{{Cruikshank} DP}, \bibinfo{editor}{{Morbidelli} A} and  \bibinfo{editor}{{Dotson} R}, (Eds.), \bibinfo{booktitle}{The Solar System Beyond Neptune},  \bibinfo{pages}{43--57}.

\bibtype{Article}%
\bibitem[{Gladman} et al.(2009)]{Gladman2009}
\bibinfo{author}{{Gladman} B}, \bibinfo{author}{{Kavelaars} J}, \bibinfo{author}{{Petit} JM}, \bibinfo{author}{{Ashby} MLN}, \bibinfo{author}{{Parker} J}, \bibinfo{author}{{Coffey} J}, \bibinfo{author}{{Jones} RL}, \bibinfo{author}{{Rousselot} P} and  \bibinfo{author}{{Mousis} O} (\bibinfo{year}{2009}), \bibinfo{month}{Jun.}
\bibinfo{title}{{Discovery of the First Retrograde Transneptunian Object}}.
\bibinfo{journal}{{\em \apjl}} \bibinfo{volume}{697} (\bibinfo{number}{2}): \bibinfo{pages}{L91--L94}. \bibinfo{doi}{\doi{10.1088/0004-637X/697/2/L91}}.

\bibtype{Article}%
\bibitem[{Gladman} et al.(2012)]{Gladman2012}
\bibinfo{author}{{Gladman} B}, \bibinfo{author}{{Lawler} SM}, \bibinfo{author}{{Petit} JM}, \bibinfo{author}{{Kavelaars} J}, \bibinfo{author}{{Jones} RL}, \bibinfo{author}{{Parker} JW}, \bibinfo{author}{{Van Laerhoven} C}, \bibinfo{author}{{Nicholson} P}, \bibinfo{author}{{Rousselot} P}, \bibinfo{author}{{Bieryla} A} and  \bibinfo{author}{{Ashby} MLN} (\bibinfo{year}{2012}), \bibinfo{month}{Jul.}
\bibinfo{title}{{The Resonant Trans-Neptunian Populations}}.
\bibinfo{journal}{{\em \aj}} \bibinfo{volume}{144} (\bibinfo{number}{1}), \bibinfo{eid}{23}. \bibinfo{doi}{\doi{10.1088/0004-6256/144/1/23}}.
\eprint{1205.7065}.

\bibtype{Article}%
\bibitem[{Hahn} and {Malhotra}(2005)]{Hahn2005}
\bibinfo{author}{{Hahn} JM} and  \bibinfo{author}{{Malhotra} R} (\bibinfo{year}{2005}), \bibinfo{month}{Nov.}
\bibinfo{title}{{Neptune's Migration into a Stirred-Up Kuiper Belt: A Detailed Comparison of Simulations to Observations}}.
\bibinfo{journal}{{\em \aj}} \bibinfo{volume}{130} (\bibinfo{number}{5}): \bibinfo{pages}{2392--2414}. \bibinfo{doi}{\doi{10.1086/452638}}.
\eprint{astro-ph/0507319}.

\bibtype{Article}%
\bibitem[{Holman} et al.(2018)]{Holman2018}
\bibinfo{author}{{Holman} MJ}, \bibinfo{author}{{Payne} MJ}, \bibinfo{author}{{Blankley} P}, \bibinfo{author}{{Janssen} R} and  \bibinfo{author}{{Kuindersma} S} (\bibinfo{year}{2018}), \bibinfo{month}{Sep.}
\bibinfo{title}{{HelioLinC: A Novel Approach to the Minor Planet Linking Problem}}.
\bibinfo{journal}{{\em \aj}} \bibinfo{volume}{156} (\bibinfo{number}{3}), \bibinfo{eid}{135}. \bibinfo{doi}{\doi{10.3847/1538-3881/aad69a}}.

\bibtype{Article}%
\bibitem[{Huang} and {Gladman}(2024)]{Huang2024}
\bibinfo{author}{{Huang} Y} and  \bibinfo{author}{{Gladman} B} (\bibinfo{year}{2024}), \bibinfo{month}{Feb.}
\bibinfo{title}{{Primordial Orbital Alignment of Sednoids}}.
\bibinfo{journal}{{\em \apjl}} \bibinfo{volume}{962} (\bibinfo{number}{2}), \bibinfo{eid}{L33}. \bibinfo{doi}{\doi{10.3847/2041-8213/ad2686}}.
\eprint{2310.20614}.

\bibtype{Article}%
\bibitem[{Huang} et al.(2022)]{Huang2022}
\bibinfo{author}{{Huang} Y}, \bibinfo{author}{{Gladman} B}, \bibinfo{author}{{Beaudoin} M} and  \bibinfo{author}{{Zhang} K} (\bibinfo{year}{2022}), \bibinfo{month}{Oct.}
\bibinfo{title}{{A Rogue Planet Helps to Populate the Distant Kuiper Belt}}.
\bibinfo{journal}{{\em \apjl}} \bibinfo{volume}{938} (\bibinfo{number}{2}), \bibinfo{eid}{L23}. \bibinfo{doi}{\doi{10.3847/2041-8213/ac9480}}.
\eprint{2209.09399}.

\bibtype{Article}%
\bibitem[{Hughes} et al.(2018)]{Hughes2018}
\bibinfo{author}{{Hughes} AM}, \bibinfo{author}{{Duch{\^e}ne} G} and  \bibinfo{author}{{Matthews} BC} (\bibinfo{year}{2018}), \bibinfo{month}{Sep.}
\bibinfo{title}{{Debris Disks: Structure, Composition, and Variability}}.
\bibinfo{journal}{{\em \araa}} \bibinfo{volume}{56}: \bibinfo{pages}{541--591}. \bibinfo{doi}{\doi{10.1146/annurev-astro-081817-052035}}.
\eprint{1802.04313}.

\bibtype{Article}%
\bibitem[{Jewitt} and {Luu}(1993)]{Jewitt1993}
\bibinfo{author}{{Jewitt} D} and  \bibinfo{author}{{Luu} J} (\bibinfo{year}{1993}), \bibinfo{month}{Apr.}
\bibinfo{title}{{Discovery of the candidate Kuiper belt object 1992 QB$_{1}$}}.
\bibinfo{journal}{{\em \nat}} \bibinfo{volume}{362} (\bibinfo{number}{6422}): \bibinfo{pages}{730--732}. \bibinfo{doi}{\doi{10.1038/362730a0}}.

\bibtype{Article}%
\bibitem[{Jewitt} and {Luu}(2004)]{Jewitt2004}
\bibinfo{author}{{Jewitt} DC} and  \bibinfo{author}{{Luu} J} (\bibinfo{year}{2004}), \bibinfo{month}{Dec.}
\bibinfo{title}{{Crystalline water ice on the Kuiper belt object (50000) Quaoar}}.
\bibinfo{journal}{{\em \nat}} \bibinfo{volume}{432} (\bibinfo{number}{7018}): \bibinfo{pages}{731--733}. \bibinfo{doi}{\doi{10.1038/nature03111}}.

\bibtype{incollection}%
\bibitem[{Kavelaars} et al.(2008)]{kavelaars2008}
\bibinfo{author}{{Kavelaars} J}, \bibinfo{author}{{Jones} L}, \bibinfo{author}{{Gladman} B}, \bibinfo{author}{{Parker} JW} and  \bibinfo{author}{{Petit} JM} (\bibinfo{year}{2008}), \bibinfo{title}{{The Orbital and Spatial Distribution of the Kuiper Belt}}, \bibinfo{editor}{{Barucci} MA}, \bibinfo{editor}{{Boehnhardt} H}, \bibinfo{editor}{{Cruikshank} DP}, \bibinfo{editor}{{Morbidelli} A} and  \bibinfo{editor}{{Dotson} R}, (Eds.), \bibinfo{booktitle}{The Solar System Beyond Neptune},  \bibinfo{pages}{59--69}.

\bibtype{incollection}%
\bibitem[{Kavelaars} et al.(2020)]{Kavelaars2020}
\bibinfo{author}{{Kavelaars} JJ}, \bibinfo{author}{{Lawler} SM}, \bibinfo{author}{{Bannister} MT} and  \bibinfo{author}{{Shankman} C} (\bibinfo{year}{2020}), \bibinfo{title}{{Perspectives on the distribution of orbits of distant Trans-Neptunian objects}}, \bibinfo{editor}{{Prialnik} D}, \bibinfo{editor}{{Barucci} MA} and  \bibinfo{editor}{{Young} L}, (Eds.), \bibinfo{booktitle}{The Trans-Neptunian Solar System},  \bibinfo{pages}{61--77}.

\bibtype{Article}%
\bibitem[{Kavelaars} et al.(2021)]{Kavelaars2021}
\bibinfo{author}{{Kavelaars} JJ}, \bibinfo{author}{{Petit} JM}, \bibinfo{author}{{Gladman} B}, \bibinfo{author}{{Bannister} MT}, \bibinfo{author}{{Alexandersen} M}, \bibinfo{author}{{Chen} YT}, \bibinfo{author}{{Gwyn} SDJ} and  \bibinfo{author}{{Volk} K} (\bibinfo{year}{2021}), \bibinfo{month}{Oct.}
\bibinfo{title}{{OSSOS Finds an Exponential Cutoff in the Size Distribution of the Cold Classical Kuiper Belt}}.
\bibinfo{journal}{{\em \apjl}} \bibinfo{volume}{920} (\bibinfo{number}{2}), \bibinfo{eid}{L28}. \bibinfo{doi}{\doi{10.3847/2041-8213/ac2c72}}.
\eprint{2107.06120}.

\bibtype{Article}%
\bibitem[{Kenyon} and {Bromley}(2004)]{Kenyon2004}
\bibinfo{author}{{Kenyon} SJ} and  \bibinfo{author}{{Bromley} BC} (\bibinfo{year}{2004}), \bibinfo{month}{Dec.}
\bibinfo{title}{{Stellar encounters as the origin of distant Solar System objects in highly eccentric orbits}}.
\bibinfo{journal}{{\em \nat}} \bibinfo{volume}{432} (\bibinfo{number}{7017}): \bibinfo{pages}{598--602}. \bibinfo{doi}{\doi{10.1038/nature03136}}.
\eprint{astro-ph/0412030}.

\bibtype{Article}%
\bibitem[{Kuiper}(1951)]{Kuiper1951}
\bibinfo{author}{{Kuiper} GP} (\bibinfo{year}{1951}), \bibinfo{month}{Jan.}
\bibinfo{title}{{On the Origin of the Solar System}}.
\bibinfo{journal}{{\em Proceedings of the National Academy of Science}} \bibinfo{volume}{37} (\bibinfo{number}{1}): \bibinfo{pages}{1--14}. \bibinfo{doi}{\doi{10.1073/pnas.37.1.1}}.

\bibtype{Inproceedings}%
\bibitem[{Lawler}(2014)]{Lawler2014}
\bibinfo{author}{{Lawler} SM} (\bibinfo{year}{2014}), \bibinfo{month}{Jan.}, \bibinfo{title}{{The Debiased Kuiper Belt: Our Solar System as a Debris Disk}}, \bibinfo{editor}{{Booth} M}, \bibinfo{editor}{{Matthews} BC} and  \bibinfo{editor}{{Graham} JR}, (Eds.), \bibinfo{booktitle}{Exploring the Formation and Evolution of Planetary Systems}, \bibinfo{series}{IAU Symposium}, \bibinfo{volume}{299},  \bibinfo{pages}{232--236}, \eprint{1405.3242}.

\bibtype{Article}%
\bibitem[{Lawler} and {Gladman}(2013)]{Lawler2013}
\bibinfo{author}{{Lawler} SM} and  \bibinfo{author}{{Gladman} B} (\bibinfo{year}{2013}), \bibinfo{month}{Jul.}
\bibinfo{title}{{Plutino Detection Biases, Including the Kozai Resonance}}.
\bibinfo{journal}{{\em \aj}} \bibinfo{volume}{146} (\bibinfo{number}{1}), \bibinfo{eid}{6}. \bibinfo{doi}{\doi{10.1088/0004-6256/146/1/6}}.
\eprint{1305.1662}.

\bibtype{Article}%
\bibitem[{Lawler} et al.(2018)]{Lawler2018}
\bibinfo{author}{{Lawler} SM}, \bibinfo{author}{{Kavelaars} JJ}, \bibinfo{author}{{Alexandersen} M}, \bibinfo{author}{{Bannister} MT}, \bibinfo{author}{{Gladman} B}, \bibinfo{author}{{Petit} JM} and  \bibinfo{author}{{Shankman} C} (\bibinfo{year}{2018}), \bibinfo{month}{May}.
\bibinfo{title}{{OSSOS: X. How to use a Survey Simulator: Statistical Testing of Dynamical Models Against the Real Kuiper Belt}}.
\bibinfo{journal}{{\em Frontiers in Astronomy and Space Sciences}} \bibinfo{volume}{5}, \bibinfo{eid}{14}. \bibinfo{doi}{\doi{10.3389/fspas.2018.00014}}.
\eprint{1802.00460}.

\bibtype{Article}%
\bibitem[{Lawler} et al.(2019)]{Lawler2019}
\bibinfo{author}{{Lawler} SM}, \bibinfo{author}{{Pike} RE}, \bibinfo{author}{{Kaib} N}, \bibinfo{author}{{Alexandersen} M}, \bibinfo{author}{{Bannister} MT}, \bibinfo{author}{{Chen} YT}, \bibinfo{author}{{Gladman} B}, \bibinfo{author}{{Gwyn} S}, \bibinfo{author}{{Kavelaars} JJ}, \bibinfo{author}{{Petit} JM} and  \bibinfo{author}{{Volk} K} (\bibinfo{year}{2019}), \bibinfo{month}{Jun.}
\bibinfo{title}{{OSSOS. XIII. Fossilized Resonant Dropouts Tentatively Confirm Neptune{\textquoteright}s Migration Was Grainy and Slow}}.
\bibinfo{journal}{{\em \aj}} \bibinfo{volume}{157} (\bibinfo{number}{6}), \bibinfo{eid}{253}. \bibinfo{doi}{\doi{10.3847/1538-3881/ab1c4c}}.
\eprint{1808.02618}.

\bibtype{Article}%
\bibitem[{Lawler} et al.(2022)]{Lawler2022}
\bibinfo{author}{{Lawler} SM}, \bibinfo{author}{{Boley} AC} and  \bibinfo{author}{{Rein} H} (\bibinfo{year}{2022}), \bibinfo{month}{Jan.}
\bibinfo{title}{{Visibility Predictions for Near-future Satellite Megaconstellations: Latitudes near 50{\textdegree} Will Experience the Worst Light Pollution}}.
\bibinfo{journal}{{\em \aj}} \bibinfo{volume}{163} (\bibinfo{number}{1}), \bibinfo{eid}{21}. \bibinfo{doi}{\doi{10.3847/1538-3881/ac341b}}.
\eprint{2109.04328}.

\bibtype{Article}%
\bibitem[{Lellouch} et al.(2013)]{Lellouch2013}
\bibinfo{author}{{Lellouch} E}, \bibinfo{author}{{Santos-Sanz} P}, \bibinfo{author}{{Lacerda} P}, \bibinfo{author}{{Mommert} M}, \bibinfo{author}{{Duffard} R}, \bibinfo{author}{{Ortiz} JL}, \bibinfo{author}{{M{\"u}ller} TG}, \bibinfo{author}{{Fornasier} S}, \bibinfo{author}{{Stansberry} J}, \bibinfo{author}{{Kiss} C}, \bibinfo{author}{{Vilenius} E}, \bibinfo{author}{{Mueller} M}, \bibinfo{author}{{Peixinho} N}, \bibinfo{author}{{Moreno} R}, \bibinfo{author}{{Groussin} O}, \bibinfo{author}{{Delsanti} A} and  \bibinfo{author}{{Harris} AW} (\bibinfo{year}{2013}), \bibinfo{month}{Sep.}
\bibinfo{title}{{``TNOs are Cool'': A survey of the trans-Neptunian region. IX. Thermal properties of Kuiper belt objects and Centaurs from combined Herschel and Spitzer observations}}.
\bibinfo{journal}{{\em \aap}} \bibinfo{volume}{557}, \bibinfo{eid}{A60}. \bibinfo{doi}{\doi{10.1051/0004-6361/201322047}}.

\bibtype{Article}%
\bibitem[{Levison} and {Morbidelli}(2003)]{Levison2003}
\bibinfo{author}{{Levison} HF} and  \bibinfo{author}{{Morbidelli} A} (\bibinfo{year}{2003}), \bibinfo{month}{Nov.}
\bibinfo{title}{{The formation of the Kuiper belt by the outward transport of bodies during Neptune's migration}}.
\bibinfo{journal}{{\em \nat}} \bibinfo{volume}{426} (\bibinfo{number}{6965}): \bibinfo{pages}{419--421}. \bibinfo{doi}{\doi{10.1038/nature02120}}.

\bibtype{Article}%
\bibitem[{Li} et al.(2022)]{Li2022}
\bibinfo{author}{{Li} D}, \bibinfo{author}{{Mustill} AJ} and  \bibinfo{author}{{Davies} MB} (\bibinfo{year}{2022}), \bibinfo{month}{Jan.}
\bibinfo{title}{{Metal Pollution of the Solar White Dwarf by Solar System Small Bodies}}.
\bibinfo{journal}{{\em \apj}} \bibinfo{volume}{924} (\bibinfo{number}{2}), \bibinfo{eid}{61}. \bibinfo{doi}{\doi{10.3847/1538-4357/ac33a8}}.
\eprint{2110.12660}.

\bibtype{Article}%
\bibitem[{Lykawka} and {Ito}(2023)]{Lykawka2023}
\bibinfo{author}{{Lykawka} PS} and  \bibinfo{author}{{Ito} T} (\bibinfo{year}{2023}), \bibinfo{month}{Sep.}
\bibinfo{title}{{Is There an Earth-like Planet in the Distant Kuiper Belt?}}
\bibinfo{journal}{{\em \aj}} \bibinfo{volume}{166} (\bibinfo{number}{3}), \bibinfo{eid}{118}. \bibinfo{doi}{\doi{10.3847/1538-3881/aceaf0}}.
\eprint{2308.13765}.

\bibtype{Article}%
\bibitem[{Madigan} et al.(2018)]{Madigan2018}
\bibinfo{author}{{Madigan} AM}, \bibinfo{author}{{Zderic} A}, \bibinfo{author}{{McCourt} M} and  \bibinfo{author}{{Fleisig} J} (\bibinfo{year}{2018}), \bibinfo{month}{Oct.}
\bibinfo{title}{{On the Dynamics of the Inclination Instability}}.
\bibinfo{journal}{{\em \aj}} \bibinfo{volume}{156} (\bibinfo{number}{4}), \bibinfo{eid}{141}. \bibinfo{doi}{\doi{10.3847/1538-3881/aad95c}}.
\eprint{1805.03651}.

\bibtype{Article}%
\bibitem[{Malhotra}(1993)]{Malhotra1993}
\bibinfo{author}{{Malhotra} R} (\bibinfo{year}{1993}), \bibinfo{month}{Oct.}
\bibinfo{title}{{The origin of Pluto's peculiar orbit}}.
\bibinfo{journal}{{\em \nat}} \bibinfo{volume}{365} (\bibinfo{number}{6449}): \bibinfo{pages}{819--821}. \bibinfo{doi}{\doi{10.1038/365819a0}}.

\bibtype{Article}%
\bibitem[{McKinnon} et al.(2020)]{McKinnon2020}
\bibinfo{author}{{McKinnon} WB}, \bibinfo{author}{{Richardson} DC}, \bibinfo{author}{{Marohnic} JC}, \bibinfo{author}{{Keane} JT}, \bibinfo{author}{{Grundy} WM}, \bibinfo{author}{{Hamilton} DP}, \bibinfo{author}{{Nesvorn{\'y}} D}, \bibinfo{author}{{Umurhan} OM}, \bibinfo{author}{{Lauer} TR}, \bibinfo{author}{{Singer} KN}, \bibinfo{author}{{Stern} SA}, \bibinfo{author}{{Weaver} HA}, \bibinfo{author}{{Spencer} JR}, \bibinfo{author}{{Buie} MW}, \bibinfo{author}{{Moore} JM}, \bibinfo{author}{{Kavelaars} JJ}, \bibinfo{author}{{Lisse} CM}, \bibinfo{author}{{Mao} X}, \bibinfo{author}{{Parker} AH}, \bibinfo{author}{{Porter} SB}, \bibinfo{author}{{Showalter} MR}, \bibinfo{author}{{Olkin} CB}, \bibinfo{author}{{Cruikshank} DP}, \bibinfo{author}{{Elliott} HA}, \bibinfo{author}{{Gladstone} GR}, \bibinfo{author}{{Parker} JW}, \bibinfo{author}{{Verbiscer} AJ}, \bibinfo{author}{{Young} LA} and  \bibinfo{author}{{New Horizons Science Team}} (\bibinfo{year}{2020}), \bibinfo{month}{Feb.}
\bibinfo{title}{{The solar nebula origin of (486958) Arrokoth, a primordial contact binary in the Kuiper Belt}}.
\bibinfo{journal}{{\em Science}} \bibinfo{volume}{367} (\bibinfo{number}{6481}), \bibinfo{eid}{aay6620}. \bibinfo{doi}{\doi{10.1126/science.aay6620}}.
\eprint{2003.05576}.

\bibtype{Article}%
\bibitem[{Moro-Mart{\'\i}n} and {Malhotra}(2003)]{Moro-Martin2003}
\bibinfo{author}{{Moro-Mart{\'\i}n} A} and  \bibinfo{author}{{Malhotra} R} (\bibinfo{year}{2003}), \bibinfo{month}{Apr.}
\bibinfo{title}{{Dynamical Models of Kuiper Belt Dust in the Inner and Outer Solar System}}.
\bibinfo{journal}{{\em \aj}} \bibinfo{volume}{125} (\bibinfo{number}{4}): \bibinfo{pages}{2255--2265}. \bibinfo{doi}{\doi{10.1086/368237}}.
\eprint{astro-ph/0506703}.

\bibtype{Article}%
\bibitem[{M{\"u}ller} et al.(2009)]{Muller2009}
\bibinfo{author}{{M{\"u}ller} TG}, \bibinfo{author}{{Lellouch} E}, \bibinfo{author}{{B{\"o}hnhardt} H}, \bibinfo{author}{{Stansberry} J}, \bibinfo{author}{{Barucci} A}, \bibinfo{author}{{Crovisier} J}, \bibinfo{author}{{Delsanti} A}, \bibinfo{author}{{Doressoundiram} A}, \bibinfo{author}{{Dotto} E}, \bibinfo{author}{{Duffard} R}, \bibinfo{author}{{Fornasier} S}, \bibinfo{author}{{Groussin} O}, \bibinfo{author}{{Guti{\'e}rrez} PJ}, \bibinfo{author}{{Hainaut} O}, \bibinfo{author}{{Harris} AW}, \bibinfo{author}{{Hartogh} P}, \bibinfo{author}{{Hestroffer} D}, \bibinfo{author}{{Horner} J}, \bibinfo{author}{{Jewitt} D}, \bibinfo{author}{{Kidger} M}, \bibinfo{author}{{Kiss} C}, \bibinfo{author}{{Lacerda} P}, \bibinfo{author}{{Lara} L}, \bibinfo{author}{{Lim} T}, \bibinfo{author}{{Mueller} M}, \bibinfo{author}{{Moreno} R}, \bibinfo{author}{{Ortiz} JL}, \bibinfo{author}{{Rengel} M}, \bibinfo{author}{{Santos-Sanz} P}, \bibinfo{author}{{Swinyard} B}, \bibinfo{author}{{Thomas} N}, \bibinfo{author}{{Thirouin} A} and
  \bibinfo{author}{{Trilling} D} (\bibinfo{year}{2009}), \bibinfo{month}{Sep.}
\bibinfo{title}{{TNOs are Cool: A Survey of the Transneptunian Region}}.
\bibinfo{journal}{{\em Earth Moon and Planets}} \bibinfo{volume}{105} (\bibinfo{number}{2-4}): \bibinfo{pages}{209--219}. \bibinfo{doi}{\doi{10.1007/s11038-009-9307-x}}.

\bibtype{Book}%
\bibitem[{Murray} and {Dermott}(1999)]{Murray1999}
\bibinfo{author}{{Murray} CD} and  \bibinfo{author}{{Dermott} SF} (\bibinfo{year}{1999}).
\bibinfo{title}{{Solar System Dynamics}}.
\bibinfo{doi}{\doi{10.1017/CBO9781139174817}}.

\bibtype{Article}%
\bibitem[{Napier} et al.(2021)]{Napier2021}
\bibinfo{author}{{Napier} KJ}, \bibinfo{author}{{Gerdes} DW}, \bibinfo{author}{{Lin} HW}, \bibinfo{author}{{Hamilton} SJ}, \bibinfo{author}{{Bernstein} GM}, \bibinfo{author}{{Bernardinelli} PH}, \bibinfo{author}{{Abbott} TMC}, \bibinfo{author}{{Aguena} M}, \bibinfo{author}{{Annis} J}, \bibinfo{author}{{Avila} S}, \bibinfo{author}{{Bacon} D}, \bibinfo{author}{{Bertin} E}, \bibinfo{author}{{Brooks} D}, \bibinfo{author}{{Burke} DL}, \bibinfo{author}{{Carnero Rosell} A}, \bibinfo{author}{{Carrasco Kind} M}, \bibinfo{author}{{Carretero} J}, \bibinfo{author}{{Costanzi} M}, \bibinfo{author}{{da Costa} LN}, \bibinfo{author}{{De Vicente} J}, \bibinfo{author}{{Diehl} HT}, \bibinfo{author}{{Doel} P}, \bibinfo{author}{{Everett} S}, \bibinfo{author}{{Ferrero} I}, \bibinfo{author}{{Fosalba} P}, \bibinfo{author}{{Garc{\'\i}a-Bellido} J}, \bibinfo{author}{{Gruen} D}, \bibinfo{author}{{Gruendl} RA}, \bibinfo{author}{{Gutierrez} G}, \bibinfo{author}{{Hollowood} DL}, \bibinfo{author}{{Honscheid} K}, \bibinfo{author}{{Hoyle} B},
  \bibinfo{author}{{James} DJ}, \bibinfo{author}{{Kent} S}, \bibinfo{author}{{Kuehn} K}, \bibinfo{author}{{Kuropatkin} N}, \bibinfo{author}{{Maia} MAG}, \bibinfo{author}{{Menanteau} F}, \bibinfo{author}{{Miquel} R}, \bibinfo{author}{{Morgan} R}, \bibinfo{author}{{Palmese} A}, \bibinfo{author}{{Paz-Chinch{\'o}n} F}, \bibinfo{author}{{Plazas} AA}, \bibinfo{author}{{Sanchez} E}, \bibinfo{author}{{Scarpine} V}, \bibinfo{author}{{Serrano} S}, \bibinfo{author}{{Sevilla-Noarbe} I}, \bibinfo{author}{{Smith} M}, \bibinfo{author}{{Suchyta} E}, \bibinfo{author}{{Swanson} MEC}, \bibinfo{author}{{To} C}, \bibinfo{author}{{Walker} AR}, \bibinfo{author}{{Wilkinson} RD} and  \bibinfo{author}{{DES Collaboration}} (\bibinfo{year}{2021}), \bibinfo{month}{Apr.}
\bibinfo{title}{{No Evidence for Orbital Clustering in the Extreme Trans-Neptunian Objects}}.
\bibinfo{journal}{{\em \psj}} \bibinfo{volume}{2} (\bibinfo{number}{2}), \bibinfo{eid}{59}. \bibinfo{doi}{\doi{10.3847/PSJ/abe53e}}.
\eprint{2102.05601}.

\bibtype{Article}%
\bibitem[{Napier} et al.(2024)]{Napier2024}
\bibinfo{author}{{Napier} KJ}, \bibinfo{author}{{Lin} HW}, \bibinfo{author}{{Gerdes} DW}, \bibinfo{author}{{Adams} FC}, \bibinfo{author}{{Simpson} AM}, \bibinfo{author}{{Porter} MW}, \bibinfo{author}{{Weber} KG}, \bibinfo{author}{{Markwardt} L}, \bibinfo{author}{{Gowman} G}, \bibinfo{author}{{Smotherman} H}, \bibinfo{author}{{Bernardinelli} PH}, \bibinfo{author}{{Juri{\'c}} M}, \bibinfo{author}{{Connolly} AJ}, \bibinfo{author}{{Kalmbach} JB}, \bibinfo{author}{{Portillo} SKN}, \bibinfo{author}{{Trilling} DE}, \bibinfo{author}{{Strauss} R}, \bibinfo{author}{{Oldroyd} WJ}, \bibinfo{author}{{Trujillo} CA}, \bibinfo{author}{{Chandler} CO}, \bibinfo{author}{{Holman} MJ}, \bibinfo{author}{{Schlichting} HE} and  \bibinfo{author}{{McNeill} A} (\bibinfo{year}{2024}), \bibinfo{month}{Feb.}
\bibinfo{title}{{The DECam Ecliptic Exploration Project (DEEP). V. The Absolute Magnitude Distribution of the Cold Classical Kuiper Belt}}.
\bibinfo{journal}{{\em \psj}} \bibinfo{volume}{5} (\bibinfo{number}{2}), \bibinfo{eid}{50}. \bibinfo{doi}{\doi{10.3847/PSJ/ad1528}}.

\bibtype{Article}%
\bibitem[{Nesvorn{\'y}}(2015)]{Nesvorny2015}
\bibinfo{author}{{Nesvorn{\'y}} D} (\bibinfo{year}{2015}), \bibinfo{month}{Sep.}
\bibinfo{title}{{Jumping Neptune Can Explain the Kuiper Belt Kernel}}.
\bibinfo{journal}{{\em \aj}} \bibinfo{volume}{150} (\bibinfo{number}{3}), \bibinfo{eid}{68}. \bibinfo{doi}{\doi{10.1088/0004-6256/150/3/68}}.
\eprint{1506.06019}.

\bibtype{Article}%
\bibitem[{Nesvorn{\'y}} and {Vokrouhlick{\'y}}(2016)]{Nesvorny2016}
\bibinfo{author}{{Nesvorn{\'y}} D} and  \bibinfo{author}{{Vokrouhlick{\'y}} D} (\bibinfo{year}{2016}), \bibinfo{month}{Jul.}
\bibinfo{title}{{Neptune's Orbital Migration Was Grainy, Not Smooth}}.
\bibinfo{journal}{{\em \apj}} \bibinfo{volume}{825} (\bibinfo{number}{2}), \bibinfo{eid}{94}. \bibinfo{doi}{\doi{10.3847/0004-637X/825/2/94}}.
\eprint{1602.06988}.

\bibtype{Article}%
\bibitem[{Nesvorn{\'y}} et al.(2021)]{Nesvorny2021}
\bibinfo{author}{{Nesvorn{\'y}} D}, \bibinfo{author}{{Li} R}, \bibinfo{author}{{Simon} JB}, \bibinfo{author}{{Youdin} AN}, \bibinfo{author}{{Richardson} DC}, \bibinfo{author}{{Marschall} R} and  \bibinfo{author}{{Grundy} WM} (\bibinfo{year}{2021}), \bibinfo{month}{Feb.}
\bibinfo{title}{{Binary Planetesimal Formation from Gravitationally Collapsing Pebble Clouds}}.
\bibinfo{journal}{{\em \psj}} \bibinfo{volume}{2} (\bibinfo{number}{1}), \bibinfo{eid}{27}. \bibinfo{doi}{\doi{10.3847/PSJ/abd858}}.
\eprint{2011.07042}.

\bibtype{Article}%
\bibitem[{Ortiz} et al.(2017)]{Ortiz2017}
\bibinfo{author}{{Ortiz} JL}, \bibinfo{author}{{Santos-Sanz} P}, \bibinfo{author}{{Sicardy} B}, \bibinfo{author}{{Benedetti-Rossi} G}, \bibinfo{author}{{B{\'e}rard} D}, \bibinfo{author}{{Morales} N}, \bibinfo{author}{{Duffard} R}, \bibinfo{author}{{Braga-Ribas} F}, \bibinfo{author}{{Hopp} U}, \bibinfo{author}{{Ries} C}, \bibinfo{author}{{Nascimbeni} V}, \bibinfo{author}{{Marzari} F}, \bibinfo{author}{{Granata} V}, \bibinfo{author}{{P{\'a}l} A}, \bibinfo{author}{{Kiss} C}, \bibinfo{author}{{Pribulla} T}, \bibinfo{author}{{Kom{\v{z}}{\'\i}k} R}, \bibinfo{author}{{Hornoch} K}, \bibinfo{author}{{Pravec} P}, \bibinfo{author}{{Bacci} P}, \bibinfo{author}{{Maestripieri} M}, \bibinfo{author}{{Nerli} L}, \bibinfo{author}{{Mazzei} L}, \bibinfo{author}{{Bachini} M}, \bibinfo{author}{{Martinelli} F}, \bibinfo{author}{{Succi} G}, \bibinfo{author}{{Ciabattari} F}, \bibinfo{author}{{Mikuz} H}, \bibinfo{author}{{Carbognani} A}, \bibinfo{author}{{Gaehrken} B}, \bibinfo{author}{{Mottola} S}, \bibinfo{author}{{Hellmich} S},
  \bibinfo{author}{{Rommel} FL}, \bibinfo{author}{{Fern{\'a}ndez-Valenzuela} E}, \bibinfo{author}{{Campo Bagatin} A}, \bibinfo{author}{{Cikota} S}, \bibinfo{author}{{Cikota} A}, \bibinfo{author}{{Lecacheux} J}, \bibinfo{author}{{Vieira-Martins} R}, \bibinfo{author}{{Camargo} JIB}, \bibinfo{author}{{Assafin} M}, \bibinfo{author}{{Colas} F}, \bibinfo{author}{{Behrend} R}, \bibinfo{author}{{Desmars} J}, \bibinfo{author}{{Meza} E}, \bibinfo{author}{{Alvarez-Candal} A}, \bibinfo{author}{{Beisker} W}, \bibinfo{author}{{Gomes-Junior} AR}, \bibinfo{author}{{Morgado} BE}, \bibinfo{author}{{Roques} F}, \bibinfo{author}{{Vachier} F}, \bibinfo{author}{{Berthier} J}, \bibinfo{author}{{Mueller} TG}, \bibinfo{author}{{Madiedo} JM}, \bibinfo{author}{{Unsalan} O}, \bibinfo{author}{{Sonbas} E}, \bibinfo{author}{{Karaman} N}, \bibinfo{author}{{Erece} O}, \bibinfo{author}{{Koseoglu} DT}, \bibinfo{author}{{Ozisik} T}, \bibinfo{author}{{Kalkan} S}, \bibinfo{author}{{Guney} Y}, \bibinfo{author}{{Niaei} MS}, \bibinfo{author}{{Satir}
  O}, \bibinfo{author}{{Yesilyaprak} C}, \bibinfo{author}{{Puskullu} C}, \bibinfo{author}{{Kabas} A}, \bibinfo{author}{{Demircan} O}, \bibinfo{author}{{Alikakos} J}, \bibinfo{author}{{Charmandaris} V}, \bibinfo{author}{{Leto} G}, \bibinfo{author}{{Ohlert} J}, \bibinfo{author}{{Christille} JM}, \bibinfo{author}{{Szak{\'a}ts} R}, \bibinfo{author}{{Tak{\'a}csn{\'e} Farkas} A}, \bibinfo{author}{{Varga-Vereb{\'e}lyi} E}, \bibinfo{author}{{Marton} G}, \bibinfo{author}{{Marciniak} A}, \bibinfo{author}{{Bartczak} P}, \bibinfo{author}{{Santana-Ros} T}, \bibinfo{author}{{Butkiewicz-B M}}, \bibinfo{author}{{Dudzi{\'n}ski} G}, \bibinfo{author}{{Al{\'\i}-Lagoa} V}, \bibinfo{author}{{Gazeas} K}, \bibinfo{author}{{Tzouganatos} L}, \bibinfo{author}{{Paschalis} N}, \bibinfo{author}{{Tsamis} V}, \bibinfo{author}{{S{\'a}nchez-Lavega} A}, \bibinfo{author}{{P{\'e}rez-Hoyos} S}, \bibinfo{author}{{Hueso} R}, \bibinfo{author}{{Guirado} JC}, \bibinfo{author}{{Peris} V} and  \bibinfo{author}{{Iglesias-Marzoa} R}
  (\bibinfo{year}{2017}), \bibinfo{month}{Oct.}
\bibinfo{title}{{The size, shape, density and ring of the dwarf planet Haumea from a stellar occultation}}.
\bibinfo{journal}{{\em \nat}} \bibinfo{volume}{550} (\bibinfo{number}{7675}): \bibinfo{pages}{219--223}. \bibinfo{doi}{\doi{10.1038/nature24051}}.
\eprint{2006.03113}.

\bibtype{Article}%
\bibitem[{Peixinho} et al.(2003)]{Peixinho2003}
\bibinfo{author}{{Peixinho} N}, \bibinfo{author}{{Doressoundiram} A}, \bibinfo{author}{{Delsanti} A}, \bibinfo{author}{{Boehnhardt} H}, \bibinfo{author}{{Barucci} MA} and  \bibinfo{author}{{Belskaya} I} (\bibinfo{year}{2003}), \bibinfo{month}{Oct.}
\bibinfo{title}{{Reopening the TNOs color controversy: Centaurs bimodality and TNOs unimodality}}.
\bibinfo{journal}{{\em \aap}} \bibinfo{volume}{410}: \bibinfo{pages}{L29--L32}. \bibinfo{doi}{\doi{10.1051/0004-6361:20031420}}.
\eprint{astro-ph/0309428}.

\bibtype{Article}%
\bibitem[{Peixinho} et al.(2015)]{Peixinho2015}
\bibinfo{author}{{Peixinho} N}, \bibinfo{author}{{Delsanti} A} and  \bibinfo{author}{{Doressoundiram} A} (\bibinfo{year}{2015}), \bibinfo{month}{May}.
\bibinfo{title}{{Reanalyzing the visible colors of Centaurs and KBOs: what is there and what we might be missing}}.
\bibinfo{journal}{{\em \aap}} \bibinfo{volume}{577}, \bibinfo{eid}{A35}. \bibinfo{doi}{\doi{10.1051/0004-6361/201425436}}.
\eprint{1502.04145}.

\bibtype{Article}%
\bibitem[{Petit} et al.(2004)]{Petit2004}
\bibinfo{author}{{Petit} JM}, \bibinfo{author}{{Holman} M}, \bibinfo{author}{{Scholl} H}, \bibinfo{author}{{Kavelaars} J} and  \bibinfo{author}{{Gladman} B} (\bibinfo{year}{2004}), \bibinfo{month}{Jan.}
\bibinfo{title}{{A highly automated moving object detection package}}.
\bibinfo{journal}{{\em \mnras}} \bibinfo{volume}{347} (\bibinfo{number}{2}): \bibinfo{pages}{471--480}. \bibinfo{doi}{\doi{10.1111/j.1365-2966.2004.07217.x}}.

\bibtype{Article}%
\bibitem[{Petit} et al.(2011)]{Petit2011}
\bibinfo{author}{{Petit} JM}, \bibinfo{author}{{Kavelaars} JJ}, \bibinfo{author}{{Gladman} BJ}, \bibinfo{author}{{Jones} RL}, \bibinfo{author}{{Parker} JW}, \bibinfo{author}{{Van Laerhoven} C}, \bibinfo{author}{{Nicholson} P}, \bibinfo{author}{{Mars} G}, \bibinfo{author}{{Rousselot} P}, \bibinfo{author}{{Mousis} O}, \bibinfo{author}{{Marsden} B}, \bibinfo{author}{{Bieryla} A}, \bibinfo{author}{{Taylor} M}, \bibinfo{author}{{Ashby} MLN}, \bibinfo{author}{{Benavidez} P}, \bibinfo{author}{{Campo Bagatin} A} and  \bibinfo{author}{{Bernabeu} G} (\bibinfo{year}{2011}), \bibinfo{month}{Oct.}
\bibinfo{title}{{The Canada-France Ecliptic Plane Survey{\textemdash}Full Data Release: The Orbital Structure of the Kuiper Belt}}.
\bibinfo{journal}{{\em \aj}} \bibinfo{volume}{142} (\bibinfo{number}{4}), \bibinfo{eid}{131}. \bibinfo{doi}{\doi{10.1088/0004-6256/142/4/131}}.
\eprint{1108.4836}.

\bibtype{Misc}%
\bibitem[{Petit} et al.(2018)]{Petit2018}
\bibinfo{author}{{Petit} JM}, \bibinfo{author}{{Kavelaars} JJ}, \bibinfo{author}{{Gladman} B} and  \bibinfo{author}{{Alexandersen} M} (\bibinfo{year}{2018}), \bibinfo{month}{May}.
\bibinfo{title}{{OSS: OSSOS Survey Simulator}}.
\bibinfo{howpublished}{Astrophysics Source Code Library, record ascl:1805.014}.

\bibtype{Article}%
\bibitem[{Pike} and {Lawler}(2017)]{Pike2017}
\bibinfo{author}{{Pike} RE} and  \bibinfo{author}{{Lawler} SM} (\bibinfo{year}{2017}), \bibinfo{month}{Oct.}
\bibinfo{title}{{Details of Resonant Structures within a Nice Model Kuiper Belt: Predictions for High-perihelion TNO Detections}}.
\bibinfo{journal}{{\em \aj}} \bibinfo{volume}{154} (\bibinfo{number}{4}), \bibinfo{eid}{171}. \bibinfo{doi}{\doi{10.3847/1538-3881/aa8b65}}.
\eprint{1709.03699}.

\bibtype{Article}%
\bibitem[{Pike} et al.(2023)]{Pike2023}
\bibinfo{author}{{Pike} RE}, \bibinfo{author}{{Fraser} WC}, \bibinfo{author}{{Volk} K}, \bibinfo{author}{{Kavelaars} JJ}, \bibinfo{author}{{Marsset} M}, \bibinfo{author}{{Peixinho} N}, \bibinfo{author}{{Schwamb} ME}, \bibinfo{author}{{Bannister} MT}, \bibinfo{author}{{Peltier} L}, \bibinfo{author}{{Buchanan} LE}, \bibinfo{author}{{Benecchi} S} and  \bibinfo{author}{{Tan} NJ} (\bibinfo{year}{2023}), \bibinfo{month}{Oct.}
\bibinfo{title}{{Col-OSSOS: The Distribution of Surface Classes in Neptune's Resonances}}.
\bibinfo{journal}{{\em \psj}} \bibinfo{volume}{4} (\bibinfo{number}{10}), \bibinfo{eid}{200}. \bibinfo{doi}{\doi{10.3847/PSJ/ace2c2}}.

\bibtype{incollection}%
\bibitem[{Pinilla-Alonso} et al.(2020)]{Pinilla-Alonso2020}
\bibinfo{author}{{Pinilla-Alonso} N}, \bibinfo{author}{{Stansberry} JA} and  \bibinfo{author}{{Holler} BJ} (\bibinfo{year}{2020}), \bibinfo{title}{{Physical and Compositional Properties of Large TNOs: from Spitzer, to JWST}}, \bibinfo{editor}{{Prialnik} D}, \bibinfo{editor}{{Barucci} MA} and  \bibinfo{editor}{{Young} L}, (Eds.), \bibinfo{booktitle}{The Trans-Neptunian Solar System},  \bibinfo{pages}{395--412}.

\bibtype{Misc}%
\bibitem[{Pinilla-Alonso} et al.(2021)]{Pinilla-Alonso2021}
\bibinfo{author}{{Pinilla-Alonso} N}, \bibinfo{author}{{Bannister} M}, \bibinfo{author}{{Brunetto} R}, \bibinfo{author}{{Cruikshank} DP}, \bibinfo{author}{{De Pra} MN}, \bibinfo{author}{{Emery} JP}, \bibinfo{author}{{Guilbert-Lepoutre} A}, \bibinfo{author}{{Holler} BJ}, \bibinfo{author}{{Lorenzi} V}, \bibinfo{author}{{Muller} T}, \bibinfo{author}{{Peixinho} N}, \bibinfo{author}{{Pendleton} YJ}, \bibinfo{author}{{Stansberry} JA} and  \bibinfo{author}{{de Souza Feliciano} AC} (\bibinfo{year}{2021}), \bibinfo{month}{Mar.}
\bibinfo{title}{{DiSCo-TNOs: Discovering the composition of the trans-Neptunian objects, icy embryos for planet formation}}.
\bibinfo{howpublished}{JWST Proposal. Cycle 1, ID. \#2418}.

\bibtype{Article}%
\bibitem[{Pinilla-Alonso} et al.(2024)]{Pinilla-Alonso2024}
\bibinfo{author}{{Pinilla-Alonso} N}, \bibinfo{author}{{Licandro} J}, \bibinfo{author}{{Brunetto} R}, \bibinfo{author}{{Henault} E}, \bibinfo{author}{{Schambeau} C}, \bibinfo{author}{{Guilbert-Lepoutre} A}, \bibinfo{author}{{Stansberry} J}, \bibinfo{author}{{Wong} I}, \bibinfo{author}{{Lunine} JI}, \bibinfo{author}{{Holler} BJ}, \bibinfo{author}{{Emery} J}, \bibinfo{author}{{Protopapa} S}, \bibinfo{author}{{Cook} J}, \bibinfo{author}{{Hammel} HB}, \bibinfo{author}{{Villanueva} GL}, \bibinfo{author}{{Milam} SN}, \bibinfo{author}{{Cruikshank} D} and  \bibinfo{author}{{de Souza-Feliciano} AC} (\bibinfo{year}{2024}), \bibinfo{month}{Jul.}
\bibinfo{title}{{Unveiling the ice and gas nature of active centaur (2060) Chiron using the James Webb Space Telescope}}.
\bibinfo{journal}{{\em arXiv e-prints}} , \bibinfo{eid}{arXiv:2407.07761}\bibinfo{doi}{\doi{10.48550/arXiv.2407.07761}}.
\eprint{2407.07761}.

\bibtype{Article}%
\bibitem[{Poppe} et al.(2019)]{Poppe2019}
\bibinfo{author}{{Poppe} AR}, \bibinfo{author}{{Lisse} CM}, \bibinfo{author}{{Piquette} M}, \bibinfo{author}{{Zemcov} M}, \bibinfo{author}{{Hor{\'a}nyi} M}, \bibinfo{author}{{James} D}, \bibinfo{author}{{Szalay} JR}, \bibinfo{author}{{Bernardoni} E} and  \bibinfo{author}{{Stern} SA} (\bibinfo{year}{2019}), \bibinfo{month}{Aug.}
\bibinfo{title}{{Constraining the Solar System's Debris Disk with In Situ New Horizons Measurements from the Edgeworth-Kuiper Belt}}.
\bibinfo{journal}{{\em \apjl}} \bibinfo{volume}{881} (\bibinfo{number}{1}), \bibinfo{eid}{L12}. \bibinfo{doi}{\doi{10.3847/2041-8213/ab322a}}.

\bibtype{Article}%
\bibitem[{Rabinowitz} et al.(2006)]{Rabinowitz2006}
\bibinfo{author}{{Rabinowitz} DL}, \bibinfo{author}{{Barkume} K}, \bibinfo{author}{{Brown} ME}, \bibinfo{author}{{Roe} H}, \bibinfo{author}{{Schwartz} M}, \bibinfo{author}{{Tourtellotte} S} and  \bibinfo{author}{{Trujillo} C} (\bibinfo{year}{2006}), \bibinfo{month}{Mar.}
\bibinfo{title}{{Photometric Observations Constraining the Size, Shape, and Albedo of 2003 EL61, a Rapidly Rotating, Pluto-sized Object in the Kuiper Belt}}.
\bibinfo{journal}{{\em \apj}} \bibinfo{volume}{639} (\bibinfo{number}{2}): \bibinfo{pages}{1238--1251}. \bibinfo{doi}{\doi{10.1086/499575}}.
\eprint{astro-ph/0509401}.

\bibtype{Article}%
\bibitem[{Ragozzine} and {Brown}(2007)]{Ragozzine2007}
\bibinfo{author}{{Ragozzine} D} and  \bibinfo{author}{{Brown} ME} (\bibinfo{year}{2007}), \bibinfo{month}{Dec.}
\bibinfo{title}{{Candidate Members and Age Estimate of the Family of Kuiper Belt Object 2003 EL61}}.
\bibinfo{journal}{{\em \aj}} \bibinfo{volume}{134} (\bibinfo{number}{6}): \bibinfo{pages}{2160--2167}. \bibinfo{doi}{\doi{10.1086/522334}}.
\eprint{0709.0328}.

\bibtype{Article}%
\bibitem[{Ragozzine} and {Brown}(2009)]{Ragozzine2009}
\bibinfo{author}{{Ragozzine} D} and  \bibinfo{author}{{Brown} ME} (\bibinfo{year}{2009}), \bibinfo{month}{Jun.}
\bibinfo{title}{{Orbits and Masses of the Satellites of the Dwarf Planet Haumea (2003 EL61)}}.
\bibinfo{journal}{{\em \aj}} \bibinfo{volume}{137} (\bibinfo{number}{6}): \bibinfo{pages}{4766--4776}. \bibinfo{doi}{\doi{10.1088/0004-6256/137/6/4766}}.
\eprint{0903.4213}.

\bibtype{Article}%
\bibitem[{Rein} and {Tamayo}(2015)]{Rein2015}
\bibinfo{author}{{Rein} H} and  \bibinfo{author}{{Tamayo} D} (\bibinfo{year}{2015}), \bibinfo{month}{Sep.}
\bibinfo{title}{{WHFAST: a fast and unbiased implementation of a symplectic Wisdom-Holman integrator for long-term gravitational simulations}}.
\bibinfo{journal}{{\em \mnras}} \bibinfo{volume}{452} (\bibinfo{number}{1}): \bibinfo{pages}{376--388}. \bibinfo{doi}{\doi{10.1093/mnras/stv1257}}.
\eprint{1506.01084}.

\bibtype{Article}%
\bibitem[{Rice} and {Laughlin}(2020)]{Rice2020}
\bibinfo{author}{{Rice} M} and  \bibinfo{author}{{Laughlin} G} (\bibinfo{year}{2020}), \bibinfo{month}{Dec.}
\bibinfo{title}{{Exploring Trans-Neptunian Space with TESS: A Targeted Shift-stacking Search for Planet Nine and Distant TNOs in the Galactic Plane}}.
\bibinfo{journal}{{\em \psj}} \bibinfo{volume}{1} (\bibinfo{number}{3}), \bibinfo{eid}{81}. \bibinfo{doi}{\doi{10.3847/PSJ/abc42c}}.
\eprint{2010.13791}.

\bibtype{Article}%
\bibitem[{Robinson} et al.(2020)]{Robinson2020}
\bibinfo{author}{{Robinson} JE}, \bibinfo{author}{{Fraser} WC}, \bibinfo{author}{{Fitzsimmons} A} and  \bibinfo{author}{{Lacerda} P} (\bibinfo{year}{2020}), \bibinfo{month}{Nov.}
\bibinfo{title}{{Investigating gravitational collapse of a pebble cloud to form transneptunian binaries}}.
\bibinfo{journal}{{\em \aap}} \bibinfo{volume}{643}, \bibinfo{eid}{A55}. \bibinfo{doi}{\doi{10.1051/0004-6361/202037456}}.
\eprint{2008.04207}.

\bibtype{Article}%
\bibitem[{Showalter} and {Hamilton}(2015)]{Showalter2015}
\bibinfo{author}{{Showalter} MR} and  \bibinfo{author}{{Hamilton} DP} (\bibinfo{year}{2015}), \bibinfo{month}{Jun.}
\bibinfo{title}{{Resonant interactions and chaotic rotation of Pluto's small moons}}.
\bibinfo{journal}{{\em \nat}} \bibinfo{volume}{522} (\bibinfo{number}{7554}): \bibinfo{pages}{45--49}. \bibinfo{doi}{\doi{10.1038/nature14469}}.

\bibtype{Article}%
\bibitem[{Smotherman} et al.(2024)]{Smotherman2024}
\bibinfo{author}{{Smotherman} H}, \bibinfo{author}{{Bernardinelli} PH}, \bibinfo{author}{{Portillo} SKN}, \bibinfo{author}{{Connolly} AJ}, \bibinfo{author}{{Kalmbach} JB}, \bibinfo{author}{{Stetzler} S}, \bibinfo{author}{{Juri{\'c}} M}, \bibinfo{author}{{Bekte{\v{s}}evi{\'c}} D}, \bibinfo{author}{{Langford} Z}, \bibinfo{author}{{Adams} FC}, \bibinfo{author}{{Oldroyd} WJ}, \bibinfo{author}{{Holman} MJ}, \bibinfo{author}{{Chandler} CO}, \bibinfo{author}{{Fuentes} C}, \bibinfo{author}{{Gerdes} DW}, \bibinfo{author}{{Lin} HW}, \bibinfo{author}{{Markwardt} L}, \bibinfo{author}{{McNeill} A}, \bibinfo{author}{{Mommert} M}, \bibinfo{author}{{Napier} KJ}, \bibinfo{author}{{Payne} MJ}, \bibinfo{author}{{Ragozzine} D}, \bibinfo{author}{{Rivkin} AS}, \bibinfo{author}{{Schlichting} H}, \bibinfo{author}{{Sheppard} SS}, \bibinfo{author}{{Strauss} R}, \bibinfo{author}{{Trilling} DE} and  \bibinfo{author}{{Trujillo} CA} (\bibinfo{year}{2024}), \bibinfo{month}{Mar.}
\bibinfo{title}{{The DECam Ecliptic Exploration Project (DEEP). VI. First Multiyear Observations of Trans-Neptunian Objects}}.
\bibinfo{journal}{{\em \aj}} \bibinfo{volume}{167} (\bibinfo{number}{3}), \bibinfo{eid}{136}. \bibinfo{doi}{\doi{10.3847/1538-3881/ad1524}}.

\bibtype{Article}%
\bibitem[{Souami} and {Souchay}(2012)]{Souami2012}
\bibinfo{author}{{Souami} D} and  \bibinfo{author}{{Souchay} J} (\bibinfo{year}{2012}), \bibinfo{month}{Jul.}
\bibinfo{title}{{The solar system's invariable plane}}.
\bibinfo{journal}{{\em \aap}} \bibinfo{volume}{543}, \bibinfo{eid}{A133}. \bibinfo{doi}{\doi{10.1051/0004-6361/201219011}}.

\bibtype{Article}%
\bibitem[{Steinpilz} et al.(2020)]{Steinpilz2020}
\bibinfo{author}{{Steinpilz} T}, \bibinfo{author}{{Joeris} K}, \bibinfo{author}{{Jungmann} F}, \bibinfo{author}{{Wolf} D}, \bibinfo{author}{{Brendel} L}, \bibinfo{author}{{Teiser} J}, \bibinfo{author}{{Shinbrot} T} and  \bibinfo{author}{{Wurm} G} (\bibinfo{year}{2020}), \bibinfo{month}{Jan.}
\bibinfo{title}{{Electrical charging overcomes the bouncing barrier in planet formation}}.
\bibinfo{journal}{{\em Nature Physics}} \bibinfo{volume}{16} (\bibinfo{number}{2}): \bibinfo{pages}{225--229}. \bibinfo{doi}{\doi{10.1038/s41567-019-0728-9}}.

\bibtype{Article}%
\bibitem[{Stern} et al.(2019)]{Stern2019}
\bibinfo{author}{{Stern} SA}, \bibinfo{author}{{Weaver} HA}, \bibinfo{author}{{Spencer} JR}, \bibinfo{author}{{Olkin} CB}, \bibinfo{author}{{Gladstone} GR}, \bibinfo{author}{{Grundy} WM}, \bibinfo{author}{{Moore} JM}, \bibinfo{author}{{Cruikshank} DP}, \bibinfo{author}{{Elliott} HA}, \bibinfo{author}{{McKinnon} WB} and  \bibinfo{author}{et~al.} (\bibinfo{year}{2019}), \bibinfo{month}{May}.
\bibinfo{title}{{Initial results from the New Horizons exploration of 2014 MU$_{69}$, a small Kuiper Belt object}}.
\bibinfo{journal}{{\em Science}} \bibinfo{volume}{364} (\bibinfo{number}{6441}), \bibinfo{eid}{aaw9771}. \bibinfo{doi}{\doi{10.1126/science.aaw9771}}.
\eprint{2004.01017}.

\bibtype{incollection}%
\bibitem[{Stern} et al.(2020)]{Stern2020}
\bibinfo{author}{{Stern} SA}, \bibinfo{author}{{Spencer} JR}, \bibinfo{author}{{Verbiscer} A}, \bibinfo{author}{{Elliott} HE} and  \bibinfo{author}{{Porter} SP} (\bibinfo{year}{2020}), \bibinfo{title}{{Initial results from the exploration of the Kuiper belt by New Horizons}}, \bibinfo{editor}{{Prialnik} D}, \bibinfo{editor}{{Barucci} MA} and  \bibinfo{editor}{{Young} L}, (Eds.), \bibinfo{booktitle}{The Trans-Neptunian Solar System},  \bibinfo{pages}{379--394}.

\bibtype{Article}%
\bibitem[{Tegler} et al.(2003)]{Tegler2003}
\bibinfo{author}{{Tegler} SC}, \bibinfo{author}{{Romanishin} W} and  \bibinfo{author}{{Consolmagno} GJ} (\bibinfo{year}{2003}), \bibinfo{month}{Dec.}
\bibinfo{title}{{Color Patterns in the Kuiper Belt: A Possible Primordial Origin}}.
\bibinfo{journal}{{\em \apjl}} \bibinfo{volume}{599} (\bibinfo{number}{1}): \bibinfo{pages}{L49--L52}. \bibinfo{doi}{\doi{10.1086/381076}}.

\bibtype{Article}%
\bibitem[{Thirouin} and {Sheppard}(2022)]{Thirouin2022}
\bibinfo{author}{{Thirouin} A} and  \bibinfo{author}{{Sheppard} SS} (\bibinfo{year}{2022}), \bibinfo{month}{Jul.}
\bibinfo{title}{{Lightcurves and Rotations of Trans-Neptunian Objects in the 2:1 Mean Motion Resonance with Neptune}}.
\bibinfo{journal}{{\em \psj}} \bibinfo{volume}{3} (\bibinfo{number}{7}), \bibinfo{eid}{178}. \bibinfo{doi}{\doi{10.3847/PSJ/ac7ab8}}.
\eprint{2206.09949}.

\bibtype{incollection}%
\bibitem[{Tombaugh}(1997)]{Tombaugh1997}
\bibinfo{author}{{Tombaugh} CW} (\bibinfo{year}{1997}), \bibinfo{title}{{The Discovery of the Ninth Planet, Pluto, in 1930}}, \bibinfo{editor}{{Stern} SA} and  \bibinfo{editor}{{Tholen} DJ}, (Eds.), \bibinfo{booktitle}{Pluto and Charon}, pp.~\bibinfo{pages}{xv}.

\bibtype{Article}%
\bibitem[{Trilling} et al.(2024)]{Trilling2024}
\bibinfo{author}{{Trilling} DE}, \bibinfo{author}{{Gerdes} DW}, \bibinfo{author}{{Juri{\'c}} M}, \bibinfo{author}{{Trujillo} CA}, \bibinfo{author}{{Bernardinelli} PH}, \bibinfo{author}{{Napier} KJ}, \bibinfo{author}{{Smotherman} H}, \bibinfo{author}{{Strauss} R}, \bibinfo{author}{{Fuentes} C}, \bibinfo{author}{{Holman} MJ}, \bibinfo{author}{{Lin} HW}, \bibinfo{author}{{Markwardt} L}, \bibinfo{author}{{McNeill} A}, \bibinfo{author}{{Mommert} M}, \bibinfo{author}{{Oldroyd} WJ}, \bibinfo{author}{{Payne} MJ}, \bibinfo{author}{{Ragozzine} D}, \bibinfo{author}{{Rivkin} AS}, \bibinfo{author}{{Schlichting} H}, \bibinfo{author}{{Sheppard} SS}, \bibinfo{author}{{Adams} FC} and  \bibinfo{author}{{Chandler} CO} (\bibinfo{year}{2024}), \bibinfo{month}{Mar.}
\bibinfo{title}{{The DECam Ecliptic Exploration Project (DEEP). I. Survey Description, Science Questions, and Technical Demonstration}}.
\bibinfo{journal}{{\em \aj}} \bibinfo{volume}{167} (\bibinfo{number}{3}), \bibinfo{eid}{132}. \bibinfo{doi}{\doi{10.3847/1538-3881/ad1529}}.

\bibtype{Article}%
\bibitem[{Trujillo} et al.(2007)]{Trujillo2007}
\bibinfo{author}{{Trujillo} CA}, \bibinfo{author}{{Brown} ME}, \bibinfo{author}{{Barkume} KM}, \bibinfo{author}{{Schaller} EL} and  \bibinfo{author}{{Rabinowitz} DL} (\bibinfo{year}{2007}), \bibinfo{month}{Feb.}
\bibinfo{title}{{The Surface of 2003 EL$_{61}$ in the Near-Infrared}}.
\bibinfo{journal}{{\em \apj}} \bibinfo{volume}{655} (\bibinfo{number}{2}): \bibinfo{pages}{1172--1178}. \bibinfo{doi}{\doi{10.1086/509861}}.
\eprint{astro-ph/0601618}.

\bibtype{Article}%
\bibitem[{Tsiganis} et al.(2005)]{Tsiganis2005}
\bibinfo{author}{{Tsiganis} K}, \bibinfo{author}{{Gomes} R}, \bibinfo{author}{{Morbidelli} A} and  \bibinfo{author}{{Levison} HF} (\bibinfo{year}{2005}), \bibinfo{month}{May}.
\bibinfo{title}{{Origin of the orbital architecture of the giant planets of the Solar System}}.
\bibinfo{journal}{{\em \nat}} \bibinfo{volume}{435} (\bibinfo{number}{7041}): \bibinfo{pages}{459--461}. \bibinfo{doi}{\doi{10.1038/nature03539}}.

\bibtype{Inproceedings}%
\bibitem[{Vera C. Rubin Observatory LSST Solar System Science Collaboration} et al.(2021)]{VeraC.RubinObservatoryLSSTSolarSystemScienceCollaboration2021}
\bibinfo{author}{{Vera C. Rubin Observatory LSST Solar System Science Collaboration}}, \bibinfo{author}{{Jones} RL}, \bibinfo{author}{{Bannister} MT}, \bibinfo{author}{{Bolin} BT}, \bibinfo{author}{{Chandler} CO}, \bibinfo{author}{{Chesley} SR}, \bibinfo{author}{{Eggl} S}, \bibinfo{author}{{Greenstreet} S}, \bibinfo{author}{{Holt} TR}, \bibinfo{author}{{Hsieh} HH}, \bibinfo{author}{{Ivezic} Z}, \bibinfo{author}{{Juric} M}, \bibinfo{author}{{Kelley} MSP}, \bibinfo{author}{{Knight} MM}, \bibinfo{author}{{Malhotra} R}, \bibinfo{author}{{Oldroyd} WJ}, \bibinfo{author}{{Sarid} G}, \bibinfo{author}{{Schwamb} ME}, \bibinfo{author}{{Snodgrass} C}, \bibinfo{author}{{Solontoi} M} and  \bibinfo{author}{{Trilling} DE} (\bibinfo{year}{2021}), \bibinfo{month}{May}, \bibinfo{title}{{The Scientific Impact of the Vera C. Rubin Observatory's Legacy Survey of Space and Time (LSST) for Solar System Science}}, \bibinfo{booktitle}{Bulletin of the American Astronomical Society}, \bibinfo{volume}{53}, pp. \bibinfo{pages}{236}.

\bibtype{Article}%
\bibitem[{Veras} et al.(2014)]{Veras2014}
\bibinfo{author}{{Veras} D}, \bibinfo{author}{{Evans} NW}, \bibinfo{author}{{Wyatt} MC} and  \bibinfo{author}{{Tout} CA} (\bibinfo{year}{2014}), \bibinfo{month}{Jan.}
\bibinfo{title}{{The great escape - III. Placing post-main-sequence evolution of planetary and binary systems in a Galactic context}}.
\bibinfo{journal}{{\em \mnras}} \bibinfo{volume}{437} (\bibinfo{number}{2}): \bibinfo{pages}{1127--1140}. \bibinfo{doi}{\doi{10.1093/mnras/stt1905}}.
\eprint{1310.1395}.

\bibtype{Article}%
\bibitem[{Vilenius} et al.(2014)]{Vilenius2014}
\bibinfo{author}{{Vilenius} E}, \bibinfo{author}{{Kiss} C}, \bibinfo{author}{{M{\"u}ller} T}, \bibinfo{author}{{Mommert} M}, \bibinfo{author}{{Santos-Sanz} P}, \bibinfo{author}{{P{\'a}l} A}, \bibinfo{author}{{Stansberry} J}, \bibinfo{author}{{Mueller} M}, \bibinfo{author}{{Peixinho} N}, \bibinfo{author}{{Lellouch} E}, \bibinfo{author}{{Fornasier} S}, \bibinfo{author}{{Delsanti} A}, \bibinfo{author}{{Thirouin} A}, \bibinfo{author}{{Ortiz} JL}, \bibinfo{author}{{Duffard} R}, \bibinfo{author}{{Perna} D} and  \bibinfo{author}{{Henry} F} (\bibinfo{year}{2014}), \bibinfo{month}{Apr.}
\bibinfo{title}{{``TNOs are Cool'': A survey of the trans-Neptunian region. X. Analysis of classical Kuiper belt objects from Herschel and Spitzer observations}}.
\bibinfo{journal}{{\em \aap}} \bibinfo{volume}{564}, \bibinfo{eid}{A35}. \bibinfo{doi}{\doi{10.1051/0004-6361/201322416}}.
\eprint{1403.6309}.

\bibtype{Article}%
\bibitem[{Volk} and {Malhotra}(2017)]{Volk2017}
\bibinfo{author}{{Volk} K} and  \bibinfo{author}{{Malhotra} R} (\bibinfo{year}{2017}), \bibinfo{month}{Aug.}
\bibinfo{title}{{The Curiously Warped Mean Plane of the Kuiper Belt}}.
\bibinfo{journal}{{\em \aj}} \bibinfo{volume}{154} (\bibinfo{number}{2}), \bibinfo{eid}{62}. \bibinfo{doi}{\doi{10.3847/1538-3881/aa79ff}}.
\eprint{1704.02444}.

\bibtype{Article}%
\bibitem[{Volk} and {Malhotra}(2024)]{Volk2024b}
\bibinfo{author}{{Volk} K} and  \bibinfo{author}{{Malhotra} R} (\bibinfo{year}{2024}), \bibinfo{month}{May}.
\bibinfo{title}{{Machine Learning Assisted Dynamical Classification of Trans-Neptunian Objects}}.
\bibinfo{journal}{{\em arXiv e-prints}} , \bibinfo{eid}{arXiv:2405.05185}\bibinfo{doi}{\doi{10.48550/arXiv.2405.05185}}.
\eprint{2405.05185}.

\bibtype{Article}%
\bibitem[{Volk} and {Van Laerhoven}(2024)]{Volk2024}
\bibinfo{author}{{Volk} K} and  \bibinfo{author}{{Van Laerhoven} C} (\bibinfo{year}{2024}), \bibinfo{month}{Jan.}
\bibinfo{title}{{Dynamical Classifications of Multi-opposition TNOs as of 2023 December}}.
\bibinfo{journal}{{\em Research Notes of the American Astronomical Society}} \bibinfo{volume}{8} (\bibinfo{number}{1}), \bibinfo{eid}{36}. \bibinfo{doi}{\doi{10.3847/2515-5172/ad22d4}}.

\bibtype{Article}%
\bibitem[{Wan} et al.(2001)]{Wan2001}
\bibinfo{author}{{Wan} XS}, \bibinfo{author}{{Huang} TY} and  \bibinfo{author}{{Innanen} KA} (\bibinfo{year}{2001}), \bibinfo{month}{Feb.}
\bibinfo{title}{{The 1:1 Superresonance in Pluto's Motion}}.
\bibinfo{journal}{{\em \aj}} \bibinfo{volume}{121} (\bibinfo{number}{2}): \bibinfo{pages}{1155--1162}. \bibinfo{doi}{\doi{10.1086/318733}}.

\bibtype{Article}%
\bibitem[{Williams} and {Benson}(1971)]{Williams1971}
\bibinfo{author}{{Williams} JG} and  \bibinfo{author}{{Benson} GS} (\bibinfo{year}{1971}), \bibinfo{month}{Mar.}
\bibinfo{title}{{Resonances in the Neptune-Pluto System}}.
\bibinfo{journal}{{\em \aj}} \bibinfo{volume}{76}: \bibinfo{pages}{167}. \bibinfo{doi}{\doi{10.1086/111100}}.

\bibtype{Article}%
\bibitem[{Winn} and {Fabrycky}(2015)]{Winn2015}
\bibinfo{author}{{Winn} JN} and  \bibinfo{author}{{Fabrycky} DC} (\bibinfo{year}{2015}), \bibinfo{month}{Aug.}
\bibinfo{title}{{The Occurrence and Architecture of Exoplanetary Systems}}.
\bibinfo{journal}{{\em \araa}} \bibinfo{volume}{53}: \bibinfo{pages}{409--447}. \bibinfo{doi}{\doi{10.1146/annurev-astro-082214-122246}}.
\eprint{1410.4199}.

\bibtype{Article}%
\bibitem[{Youdin} and {Shu}(2002)]{Youdin2002}
\bibinfo{author}{{Youdin} AN} and  \bibinfo{author}{{Shu} FH} (\bibinfo{year}{2002}), \bibinfo{month}{Nov.}
\bibinfo{title}{{Planetesimal Formation by Gravitational Instability}}.
\bibinfo{journal}{{\em \apj}} \bibinfo{volume}{580} (\bibinfo{number}{1}): \bibinfo{pages}{494--505}. \bibinfo{doi}{\doi{10.1086/343109}}.
\eprint{astro-ph/0207536}.

\end{thebibliography*}

\end{document}